\documentclass[fleqn,usenatbib]{mnras}

\usepackage{newtxtext,newtxmath}
\usepackage[T1]{fontenc}
\usepackage{ae,aecompl}

\usepackage{graphicx}	% Including figure files
\usepackage{amsmath}	% Advanced maths commands
\usepackage{amssymb}	% Extra maths symbols
\usepackage{caption}
\usepackage{subcaption}
\usepackage{color,soul}
\usepackage{ulem}
\usepackage[T1]{fontenc}
\usepackage{ae,aecompl}
\usepackage{booktabs,array,dcolumn}

\usepackage[bordercolor=white,backgroundcolor=gray!30,linecolor=black,colorinlistoftodos]{todonotes}

\usepackage{longtable}
\usepackage{lipsum} % just for dummy text- not needed for a longtable

\newcolumntype{d}{D{.}{.}{2.3}}
\newcolumntype{C}{>{\centering}p}

\title[P--L relation of RSGs]{The period--luminosity relation of red supergiants with Gaia DR2}

\author[F.W. Chatys et al.]{
Filip W. Chatys,$^{1}$$^{,2}$\thanks{E-mail: fcha8613@uni.sydney.edu.au}
Timothy R. Bedding,$^{1}$$^{,2}$
Simon J. Murphy,$^{1}$$^{,2}$
L\'aszl\'o L. Kiss$^{1}$$^{,3}$$^{,6}$
\newauthor Dougal 
Dobie$^{1}$$^{,4}$ and Jonathan E. Grindlay$^{5}$
\\
$^{1}$Sydney Institute for Astronomy, School of Physics, University of Sydney, NSW 2006, Australia\\
$^{2}$Stellar Astrophysics Centre, Department of Physics and Astronomy, Aarhus University, DK-8000 Aarhus C, Denmark\\
$^{3}$Konkoly Observatory, Research Centre for Astronomy and Earth Sciences, Hungarian Academy of Sciences, H-1121 Budapest,\\
Konkoly Thege M. ut 15-17, Hungary\\
$^{4}$CSIRO Astronomy and Space Science, P.O. Box 76, Epping, New South Wales 1710, Australia\\
$^{5}$Harvard University, Center for Astrophysics, Cambridge, MA USA\\
$^{6}$MTA CSFK Lend\"ulet Near-field Cosmology Research Group\\
}

\date{Accepted XXX. Received YYY; in original form ZZZ}

\pubyear{2015}
\begin{document}
\label{firstpage}
\pagerange{\pageref{firstpage}--\pageref{lastpage}}
\maketitle
% Abstract of the paper
\begin{abstract}
We revisit the $K$\,-band period--luminosity (P--L) relations of Galactic red supergiants using Gaia Data Release 2 parallaxes and up to 70\,yr of photometry from AAVSO and ASAS campaigns. In addition, we examine $206$ LMC red supergiants using $50$\,yr of photometric data from the Digitised Harvard Astronomical Plate Collection. 
We identified periods by computing power spectra and calculated the period--luminosity relations of our samples and compared them with the literature. Newly available data tighten the P--L relations substantially. Identified periods form two groups: one with periods of $300$--$1000$ days, corresponding to pulsations, and another with Long Secondary Periods between $1000$ and $8000$ days. Among the $48$ Galactic objects we find shorter periods in $25$ stars and long secondary periods in $23$ stars. In the LMC sample we identify $85$ and $94$ red supergiants with shorter and long secondary periods, respectively.
The P--L relation of the Galactic red supergiants is in agreement with the red supergiants in both, the Large Magellanic Cloud and the Andromeda Galaxy. We find no clear continuity between the known red giant period-luminosity sequences, and the red supergiant sequences investigated here.
%RSGs have their longer periods more dispersed than the LSPs of the compared red giant stars; Also shorter periods of both RSGs and RGs seem to be grouped in much tighter sequences than the LSPs of analysed RSGs.
\end{abstract}
% Select between one and six entries from the list of approved keywords.
% Don't make up new ones.
\begin{keywords}
stars: evolution \ -- stars: late \ -- type stars: supergiants \ --stars: pulsations \ -- solar neighbourhood
\end{keywords}

%%%%%%%%%%%%%%%%%%%%%%%%%%%%%%%%%%%%%%%%%%%%%%%%%%

%%%%%%%%%%%%%%%%% BODY OF PAPER %%%%%%%%%%%%%%%%%%

\section{Introduction}

Red supergiants (RSGs) make up some of the brightest stars in the sky, with Betelgeuse ($\alpha$ Ori) and Antares ($\alpha$ Sco) being prominent examples. RSGs are bright enough that their variability can be studied  in the Andromeda galaxy (M31) (see \citealt{Soraisam+2018}). Pulsation in RSGs is common, and they are known to follow P--L relations, which we revisit with parallaxes from Gaia Data Release 2 \citep[DR2;][]{GAIAcollab+2016,GAIAcollab+2018}.

RSGs are evolved, yet relatively young (${\leq}$ 20 M\,yr) stars. They burn helium in their cores and are very bright, i.e., \mbox{$10^5$--$10^6$\,$L$/$L$\sun}  \citep{HumphreysDavidson+1979} and moderately cool, with effective temperatures ranging from $3000$ to $5000$\,K. Most of the flux of RSGs is emitted at red and infrared wavelengths, where W Cep and $\mu$ Cep, some of the brightest Galactic RSGs, have absolute $K$\,magnitudes brighter than $-12$\,mag (see Section\:{\ref{sec:Results}} for further discussion).

The lightcurves of RSGs are either semi--periodic or irregular, which led to a suggestion that their pulsations may be stochastically excited, with a strong contribution from the convective motions (\citealt{Schwarzschild+1975}, \citealt{Dalsgaard+2001}, \citealt{Bedding+2003}, \citealt{Kiss++2006}). Changes in the circumstellar dust distribution and its composition from mass loss should also produce photometric variations in RSGs (\citealt{Meynet+2015}). The dominant variability, however, is usually attributed to radial pulsations and follows a period--luminosity (P--L) relation (\citealt{Kiss++2006}, \citealt{Pierce+2000}, \citealt{YangJiang+2011} and \citealt{GuoLi+2002}). RSGs are therefore potential ``standard candles'' for extragalactic distances (\citealt{Glass+1979}; \citealt{Feast+1980}; \citealt{Wood+1985}; \citealt{Mould+1987}). 

Another interesting property of RSGs, which they share with red giants (RGs), is the presence of long secondary periods (LSPs). These LSPs are observed in at least one third of RGs (\citealt{Wood+2000}, \citealt{Soszynski+2007}), and their origins have been debated for decades. Binarity (\citealt{Soszynski+2014}) and turnover of their giant convective cells (\citealt{Stothers+2010}) are the explanations most commonly suggested for the LSPs in RSGs, but no single mechanism has been accepted. 

With the release of Gaia DR2 parallaxes \citep{Gaia+DR2}, our aim in this work is to update our knowledge about both the Galactic and the LMC RSGs. In Sections\:{\ref{sec:MWsample}}\:and~\ref{sec:LMCsample} we describe the selection of our samples, input catalogues and the data processing. Results are shown in Section\:{\ref{sec:Results}}, where we also revisit the P--L relation of the red giants. 
%We conclude in Section~\ref{sec:conclusion}.
\section{GALACTIC RED SUPERGIANTS}
\label{sec:MWsample}

%\label{sec:ASAS-AAVSO} % used for referring to this section from elsewhere

We chose a sample of $48$ Galactic pulsating RSGs from \citet{Kiss++2006}, who measured periods using long-term visual observations from the American Association of Variable Stars Observers (AAVSO) database \footnote{\url{http://aavso.org/}}. Gaia DR2 (\citealt{Gaia+DR2}) has delivered parallaxes with uncertainties smaller than 25\% for 37 stars in this sample, up from 13 stars prior to DR2.

Our analysis of the long collections of the AAVSO photometry was supplemented by the 17\,-yr All Sky Automated Survey (ASAS) campaigns (ASAS-3 and ASAS-3N) (\citealt{Pojmanski+2004}, \citealt{ASAS_SN+2018}). Photometric measurements from ASAS used four different aperture diameters: $3$, $4$, $5$ and $6$ pixels (MAG$0$, MAG$1$, MAG$2$, MAG$3$ and MAG$4$, respectively, as per the ASAS nomenclature). We used the widest aperture MAG$4$ to capture all the flux, since contamination was not an issue for such bright objects. We analysed all available ASAS datasets (up to four available per star), each representing a different ASAS field. We found offsets in photometric measurements between both the consecutive ASAS campaigns, as well as fields within the same campaign, and in some overlapping areas magnitudes differed by as much as $0.1$\;mag. The offsets in lightcurves between AAVSO (visual estimates) and each ASAS campaign (photometry with CCD detectors) were corrected by giving the ASAS time series the same median as the AAVSO data. 

\subsection{Period analysis}
\label{sec:period} % used for referring to this section from elsewhere

From AAVSO data, we included observations of the observers who observed for more than $30$ days in total. We then binned the lightcurves into $10$, $30$ and $50$\,-d bins to: (i) minimise the effect of outliers; (ii) balance out a difference in the relative weight of the measurements between ASAS ($10$-\,yr data) and the AAVSO (at least couple of decades); (iii) make detection easier of both, shorter and longer periods ($10$, $30$\,-d bins for shorter periods and $50$\,-d bins for longer periods). The ASAS time-series were binned into \mbox{$10$\,-d} groups.
We also de-trended the AAVSO lightcurves (by subtracting a linear fit from the lightcurve) to prevent a low frequency peak from dominating the Fourier spectrum.

\begin{figure}
\includegraphics[width=\linewidth]{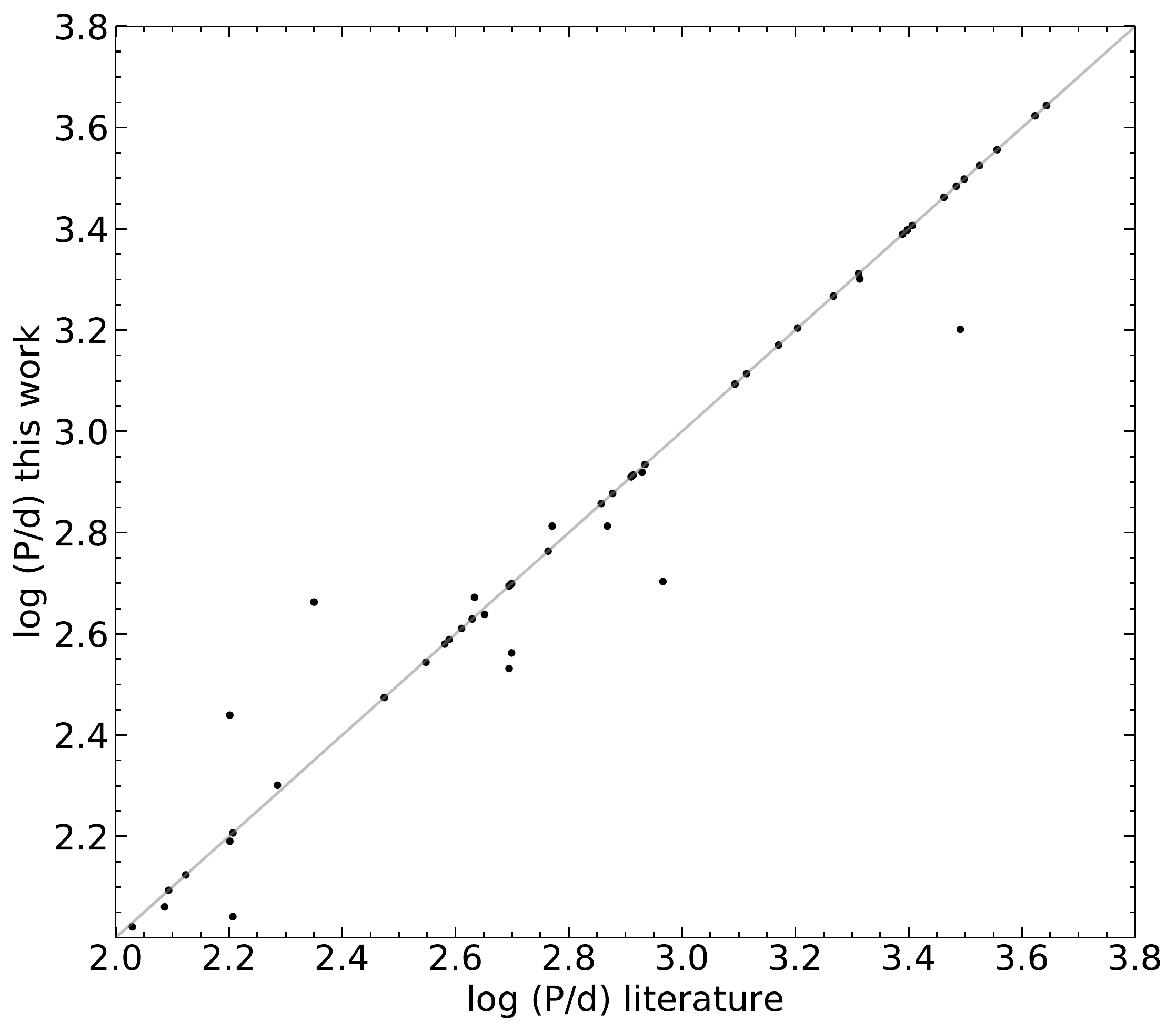}
\caption{Comparison of periods of Galactic RSGs, from this study\;(y-axis) with the literature (\citealt{Kiss++2006}, \citealt{PercyKhatu+2014})\;(x-axis). The grey line shows equality. Stars with periods, which do not agree with the literature are marked with asterisk in Table\:{\ref{table:MW Table}}.}
\label{fig:LitvsChatys}
\end{figure}

We used the Lomb-Scargle periodogram ({\citealt{Lomb}}; \citealt{Scargle}) to calculate the power spectra and identify periods. We inspected the power spectra between frequencies of $0.01$ $0.0001$\,d\textsuperscript{-1} ($100$--$10000$\,d), and searched for any distinguishable peaks.
The software {\sc Period04} \citep{Period04} was used to subtract the peak signal from the lightcurve and perform a second Fourier analysis on the residuals. When detecting periodicities, we checked against the previously identified periods (\citealt{Kiss++2006}, \citealt{Guinan+2015}, \citealt{PercyKhatu+2014}), both for consistency and to see whether there is any improvement in the findings with the recent years of data added. Figure\,{\ref{fig:LitvsChatys}} compares our measured periods with the literature ($43$ periods agreed to within $10$\%).

\begin{figure*}
\centering

\begin{subfigure}[b]{0.7\textwidth}
   \includegraphics[width=\linewidth]{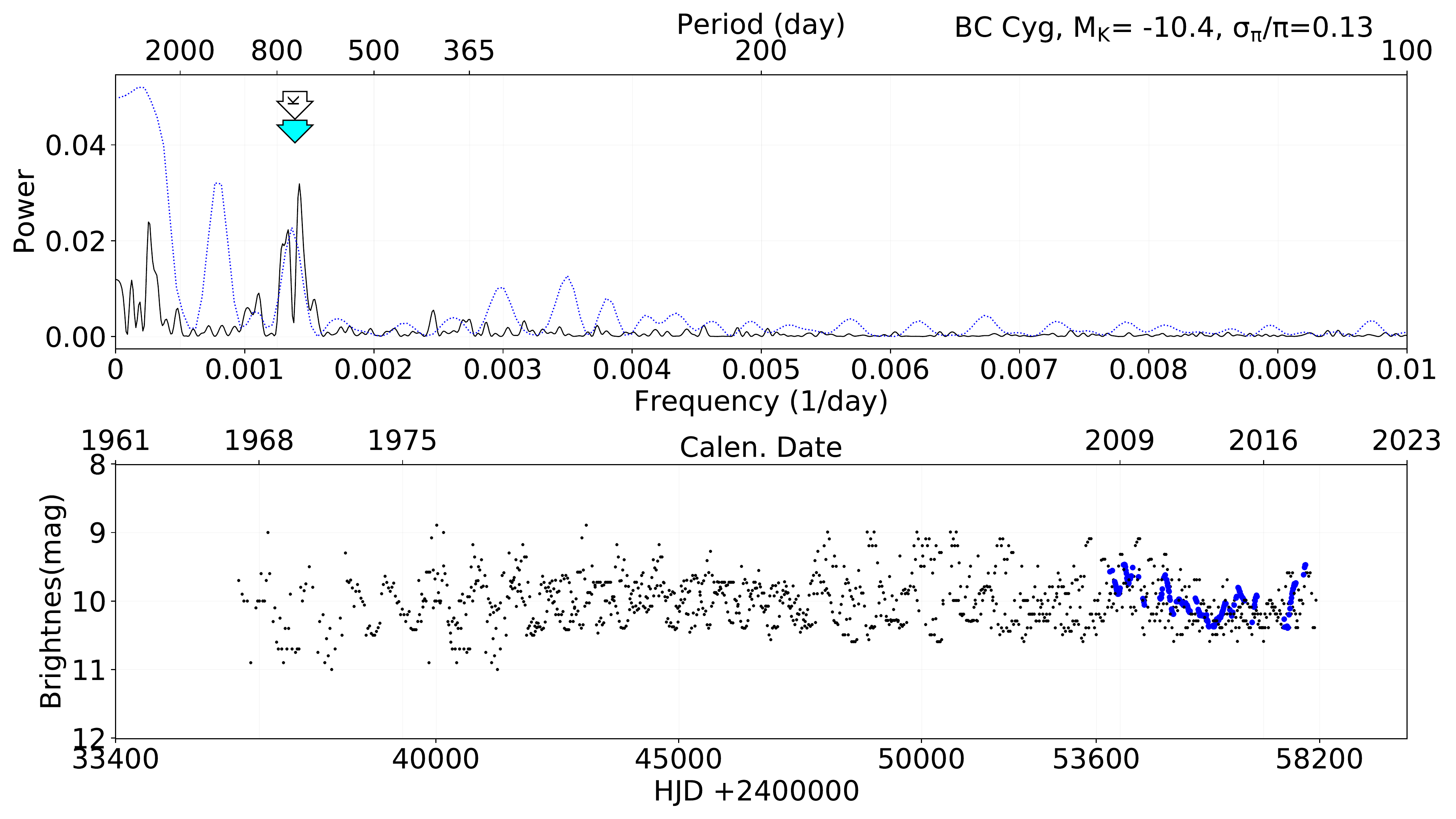}

\end{subfigure}\vspace{8mm}

\begin{subfigure}[b]{0.7\textwidth}
   \includegraphics[width=\linewidth]{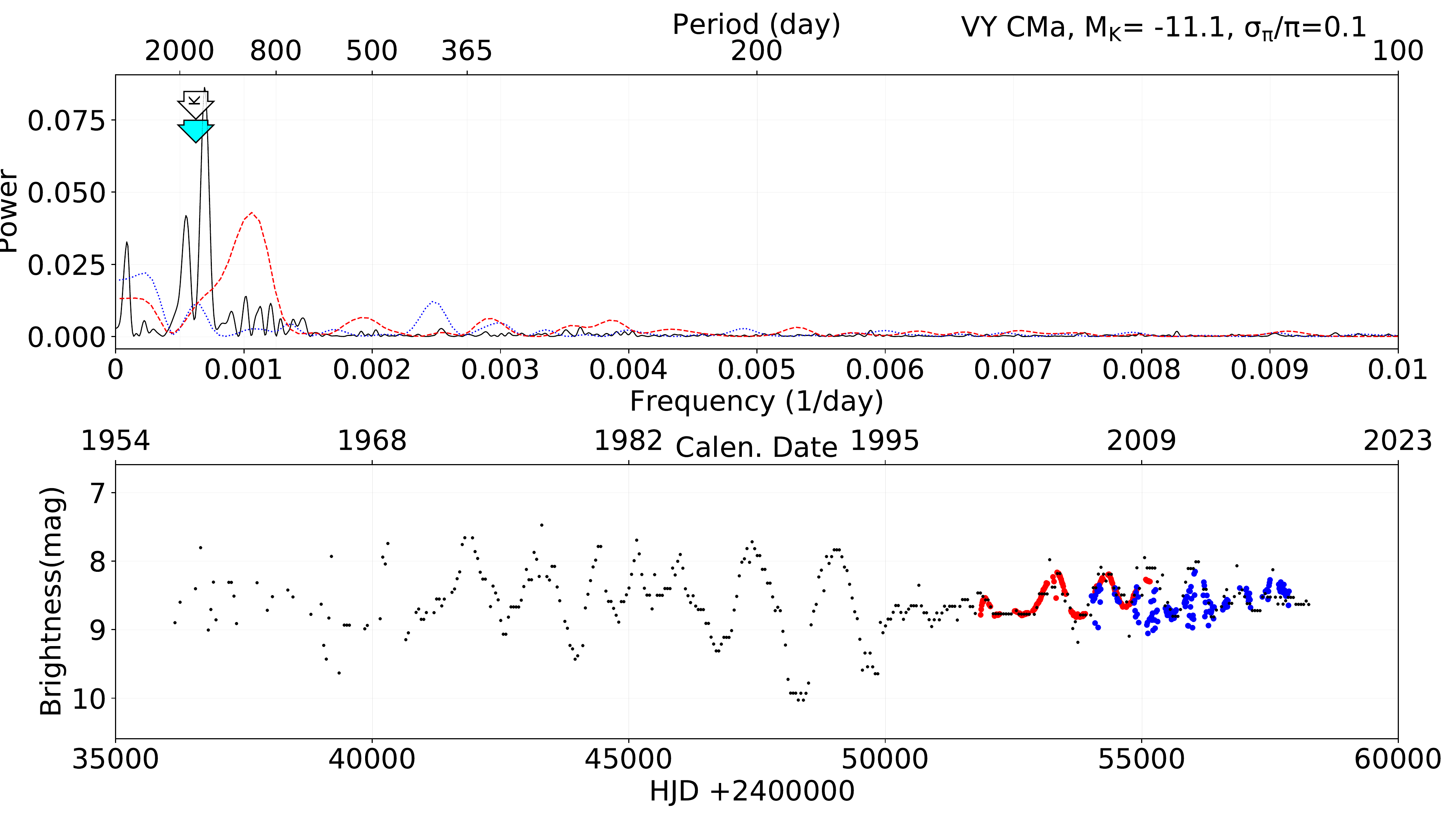}
   
  \end{subfigure}\vspace{8mm}

\begin{subfigure}[b]{0.7\textwidth}
   \includegraphics[width=\linewidth]{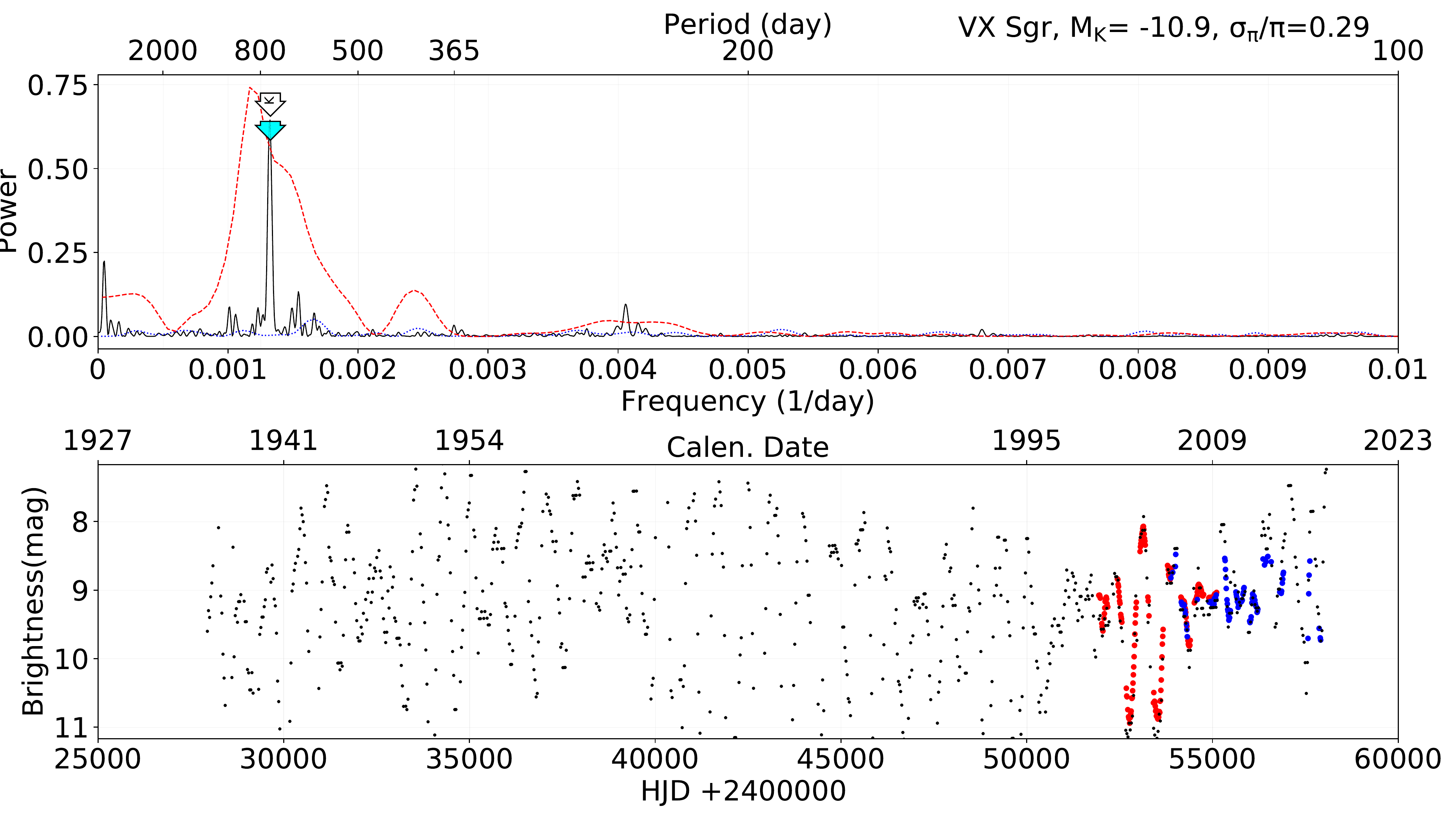}
 
  \end{subfigure}

\caption[]{Sample AAVSO ($50$\,-d bins) and ASAS ($10$\,-d bins) lightcurves with associated power spectra of three Galactic RSGs, BC~Cyg, VY~CMa and VX~Sgr. Filled arrows indicate adopted pulsational periods and white arrows with K mark periodicity published by \citet{Kiss++2006}. Black, red and blue colours indicate AAVSO, ASAS-3 and ASAS-3N data respectively (All lightcurves are available in the supplementary material).}
\label{fig:MWPS}
\end{figure*}

Representative lightcurves and power spectra are shown in Fig.\,{\ref{fig:MWPS}} for the stars BC~Cyg, VY~CMa and VX~Sgr. Notably, over half of our sample ($27$ objects) exhibit a periodicity in AAVSO data close to one year (although not the most dominant peak in the power spectrum). \citet{Kiss++2006} suggested that this effect could be caused by a seasonal variation in visibility resulting in a differential extinction of a few tenths of a magnitude. Another possibility is the Ceraski effect, described by \citet{PercyKhatu+2014}, which affects visual observers only. When they observe two stars of equal brightness that are aligned perpendicularly to the line of sight (which happens at certain times of the year), they perceive the upper star to be brighter than the one below. We omitted annual peaks from further analysis except for five stars (AZ~Cyg, $\alpha$~Ori, WY~Gem, BU~Per, $\alpha$~Her) that had these periods validated by the ASAS data.

We measured amplitudes from the height of the peak in the Fourier spectrum, which gives the semi--amplitude of the best--fitting sinusoid.
Note that amplitudes in the literature are often given as peak-to-peak values (e.g. in \citealt{YangJiang+2011}), which would be twice the values we measured. We show the ASAS amplitudes in Table\:{\ref{table:MW Table}}, which are based on CCD measurements in the V filter.

Table\:{\ref{table:MW Table}} shows the Galactic sample with stars ordered by their brightness (descending). Column $1$ shows star name, next is HD catalogue number, identified periods, amplitudes and apparent $K$ magnitude. Parallax and the associated uncertainty are in columns $10$ and $11$, respectively. Calculated absolute $K$\,magnitudes and uncertainties are shown in columns $12$, $13$ and $14$. Sources of parallaxes and $K$-band magnitudes are described in Sections\:\:{\ref{sec:DR2}} and {\ref{sec:ApMAg}}, respectively.

\begin{table*}
\caption[]{Galactic sample of RSGs. Stars are ordered by their absolute $K$ magnitudes. P1, P2 and P3 indicate the periods identified in this study with their associated amplitude values shown in amp1, amp2, amp3 columns. Parallax and the associated uncertainty are in the columns $10$ and $11$. The parallax sources are given in Section\:{\ref{sec:DR2}}. Calculated absolute $K$\,magnitudes and uncertainties are shown in columns $12$, $13$ and $14$. Asterisks mark objects for which determined periods do not agree with the literature to within $10$\%.} 
\label{table:MW Table}

\begin{tabular}{@{\extracolsep{\fill}}r r c c c c c c r c c r c c c r}
\hline
\hline

Name & HD & P1 &amp1& P2&amp2&P3 &amp3& $K$  &  $\pi$  & $\sigma_\pi$ & $M_K$ &$\sigma_{M_K}$ & $\sigma_{M_K}$   \\

 &  & (day)& (mag) & (day)& (mag) &   (day)& (mag) &  & (mas) &  (mas) & & ($+$) & ($-$)   \\
\hline

W Cep & 214369 & $10000$ & $0.24$ & --- & --- & --- & --- & $2.35$ & $0.05$ & $0.05$ & $<$$-12.65$ & --- & --- \\

$\mu$ Cep & $206936$ & $860$ & $0.06$ & $4400$ & $0.13$ & --- & --- & $-1.67$ & $0.55$ & $0.20$ & $-12.97$ & $0.67$ & $-0.98$ \\

TV Gem & $42475$ & $426$ & $0.1$ & $2550$ & $0.17$ & --- & --- & $0.91$ & $0.32$ & $0.13$ & $-11.57$ & $0.25$ & $-0.28$ \\

PZ Cas & $37536$ & $830$ & $0.21$ & --- & --- & --- & --- & $0.98$ & $0.36$ & $0.03$ & $-11.24$ & $0.38$ & $-0.46$ \\

RW Cyg & --- & $580$ & $0.19$ & --- & --- & --- & --- & $0.52$ & $0.46$ & $0.09$ & $-11.17$ & $0.39$ & $-0.47$ \\

ST Cep & $239978$ & $610$ & $0.15$ & --- & --- & --- & --- & $1.80$ & $0.26$ & $0.05$ & $-11.13$ & $0.38$ & $-0.46$ \\

VY CMa & $58061$ & $1600$ & $0.2$ & --- & --- & --- & --- & $-0.69$ & $0.83$ & $0.08$ & $-11.09$ & $0.20$ & $-0.22$ \\

SU Per & $14469$ & $470$ & $0.1$ & $3050$ & $0.24$ & --- & --- & $1.45$ & $0.32$ & $0.08$ & $-11.03$ & $0.48$ & $-0.62$ \\

VX Sgr & $165674$ & $754$ & $0.85$ & --- & --- & --- & --- & $-0.40$ & $0.79$ & $0.23$ & $-10.91$ & $0.55$ & $-0.75$ \\

NO Aur & $246070$ & --- & --- & --- & --- & --- & --- & $0.88$ & $0.45$ & $0.16$ & $-10.85$ & $0.66$ & $-0.95$ \\

S Per & $14528$ & $813$ & $0.44$ & --- & --- & --- & --- & $1.45$ & $0.41$ & $0.01$ & $-10.47$ & $0.03$ & $-0.03$ \\

CK Car* & $90382$ & $505$ & $0.14$ & --- & --- & --- & --- & $1.36$ & $0.43$ & $0.08$ & $-10.47$ & $0.37$ & $-0.45$ \\

AZ Cyg* & --- & $340$ & $0.1$ & $495$ & $0.14$ & $3350$ & $0.23$ & $1.22$ & $0.47$ & $0.08$ & $-10.42$ & $0.34$ & $-0.41$ \\

BC Cyg & --- & $720$ & $0.18$ & --- & --- & --- & --- & $0.21$ & $0.75$ & $0.10$ & $-10.41$ & $0.27$ & $-0.31$ \\

RT Car & --- & $435$ & $0.22$ & $2000$ & $0.13$ & --- & --- & $1.86$ & $0.37$ & $0.06$ & $-10.30$ & $0.33$ & $-0.38$ \\

BI Cyg & --- & $4350$ & $0.14$ & --- & --- & --- & --- & $0.59$ & $0.73$ & $0.08$ & $-10.09$ & $0.23$ & $-0.25$ \\

TZ Cas* & --- & $475$ & $0.06$ & $1590$ & $0.13$ & --- & --- & $1.89$ & $0.41$ & $0.06$ & $-10.05$ & $0.30$ & $-0.34$ \\

$\alpha$ Sco & $148478$ & --- & --- & --- & --- & --- & --- & $-3.82$ & $5.89$ & $1.00$ & $-9.96$ & $0.34$ & $-0.40$ \\

$\alpha$ Ori & $39801$ & $388$ & $0.08$ & $2050$ & $0.07$ & --- & --- & $-4.00$ & $6.55$ & $0.83$ & $-9.92$ & $0.26$ & $-0.29$ \\

IX Car & $94096$ & $408$ & $0.15$ & $4400$ & $0.15$ & --- & --- & $1.86$ & $0.45$ & $0.05$ & $-9.88$ & $0.23$ & $-0.26$ \\

XX Per & $12401$ & $3150$ & $0.03$ & --- & --- & --- & --- & $1.81$ & $0.46$ & $0.07$ & $-9.88$ & $0.31$ & $-0.36$ \\

T Per & $14142$ & $2500$ & $0.07$ & --- & --- & --- & --- & $2.66$ & $0.32$ & $0.05$ & $-9.81$ & $0.32$ & $-0.37$ \\

AO Cru & $106873$ & $3700$ & $0.12$ & --- & --- & --- & --- & $2.20$ & $0.40$ & $0.03$ & $-9.79$ & $0.16$ & $-0.17$ \\

CL Car* & $94599$ & $500$ & $0.35$ & $1400$ & $0.26$ & --- & --- & $1.68$ & $0.51$ & $0.06$ & $-9.78$ & $0.24$ & $-0.27$ \\

BU Gem & $42543$ & $2450$ & $0.19$ & --- & --- & --- & --- & $0.98$ & $0.71$ & $0.24$ & $-9.77$ & $0.63$ & $-0.90$ \\

EV Car & $89845$ & $820$ & $0.67$ & --- & --- & --- & --- & $0.90$ & $0.78$ & $0.11$ & $-9.64$ & $0.29$ & $-0.33$ \\

AD Per & $14270$ & --- & --- & --- & --- & --- & --- & $2.09$ & $0.46$ & $0.06$ & $-9.60$ & $0.27$ & $-0.30$ \\

RS Per* & $14488$ & $460$ & $0.14$ & $4200$ & $0.22$ & --- & --- & $1.68$ & $0.64$ & $0.08$ & $-9.29$ & $0.26$ & $-0.29$ \\

BO Car & $93420$ & --- & --- & --- & --- & --- & --- & $1.42$ & $0.73$ & $0.08$ & $-9.26$ & $0.23$ & $-0.25$ \\

FZ Per & $14330$ & --- & --- & --- & --- & --- & --- & $2.55$ & $0.44$ & $0.04$ & $-9.24$ & $0.19$ & $-0.21$ \\

WY Gem & $42474$ & $350$ & $0.1$ & --- & --- & --- & --- & $1.83$ & $0.63$ & $0.10$ & $-9.17$ & $0.32$ & $-0.38$ \\

PR Per & $14404$ & --- & --- & --- & --- & --- & --- & $2.34$ & $0.53$ & $0.05$ & $-9.04$ & $0.20$ & $-0.22$ \\

RV Hya & $73766$ & $195$ & $0.13$ & $950$ & $0.17$ & --- & --- & $0.52$ & $1.23$ & $0.30$ & $-9.04$ & $0.47$ & $-0.61$ \\

KK Per & $13136$ & $170$ & $0.06$ & $300$ & $0.08$ & $1850$ & $0.1$ & $2.12$ & $0.59$ & $0.05$ & $-9.03$ & $0.18$ & $-0.19$ \\

PP Per & --- & --- & --- & --- & --- & --- & --- & $2.95$ & $0.42$ & $0.05$ & $-8.93$ & $0.24$ & $-0.28$ \\

W Ind & $201866$ & $200$ & $0.5$ & --- & --- & --- & --- & $1.21$ & $1.02$ & $0.58$ & $-8.75$ & $0.98$ & $-1.83$ \\

XY Lyr & $172380$ & $115$ & $0.1$ & --- & --- & --- & --- & $-0.29$ & $2.15$ & $0.19$ & $-8.62$ & $0.18$ & $-0.20$ \\

BU Per & --- & $380$ & $0.14$ & $3600$ & $0.21$ & --- & --- & $2.26$ & $0.67$ & $0.09$ & $-8.61$ & $0.27$ & $-0.31$ \\

$\alpha$ Her* & $156014$ & $124$ & $0.04$ & $365$ & $0.06$ & $1480$ & $0.05$ & $-3.51$ & $9.91$ & $0.49$ & $-8.53$ & $0.10$ & $-0.11$ \\

AH Sco* & $155161$ & $650$ & $0.5$ & --- & --- & --- & --- & $0.31$ & $1.73$ & $0.22$ & $-8.50$ & $0.26$ & $-0.30$ \\

W Per & $237008$ & $500$ & $0.22$ & $2900$ & $0.28$ & --- & --- & $2.00$ & $0.80$ & $0.08$ & $-8.48$ & $0.21$ & $-0.23$ \\

CE Tau & $36389$ & $1300$ & $0.08$ & --- & --- & --- & --- & $-0.89$ & $3.06$ & $0.54$ & $-8.46$ & $0.35$ & $-0.42$ \\

T Cet* & $1760$ & $110$ & --- & $161$ & --- & $298$ & --- & $-0.81$ & $3.70$ & $0.47$ & $-7.97$ & $0.26$ & $-0.29$ \\

UZ Cma* & --- & $160$ & $2$ & --- & --- & --- & --- & $2.35$ & $1.06$ & $0.09$ & $-7.52$ & $0.18$ & $-0.19$ \\

Y Lyn & $58521$ & $133$ & $0.1$ & $1240$ & $0.33$ & --- & --- & $-0.46$ & $3.95$ & $0.95$ & $-7.48$ & $0.47$ & $-0.60$ \\

SS And & $218942$ & $155$ & $0.13$ & $275$ & $0.13$ & --- & --- & $0.97$ & $2.90$ & $0.89$ & $-6.72$ & $0.58$ & $-0.80$ \\

W Tri & $16682$ & $105$ & $0.07$ & $650$ & $0.06$ & --- & --- & $1.07$ & $3.31$ & $0.59$ & $-6.33$ & $0.36$ & $-0.43$ \\

IS Gem & $49380$ & --- & --- & --- & --- & --- & --- & $2.71$ & $7.64$ & $0.12$ & $-2.87$ & $0.03$ & $-0.03$ \\

\hline
\hline
\end{tabular}
\end{table*}

\subsection{Parallaxes}
\label{sec:DR2} % used for referring to this section from elsewhere

\begin{figure*}
\includegraphics[width=1\linewidth]{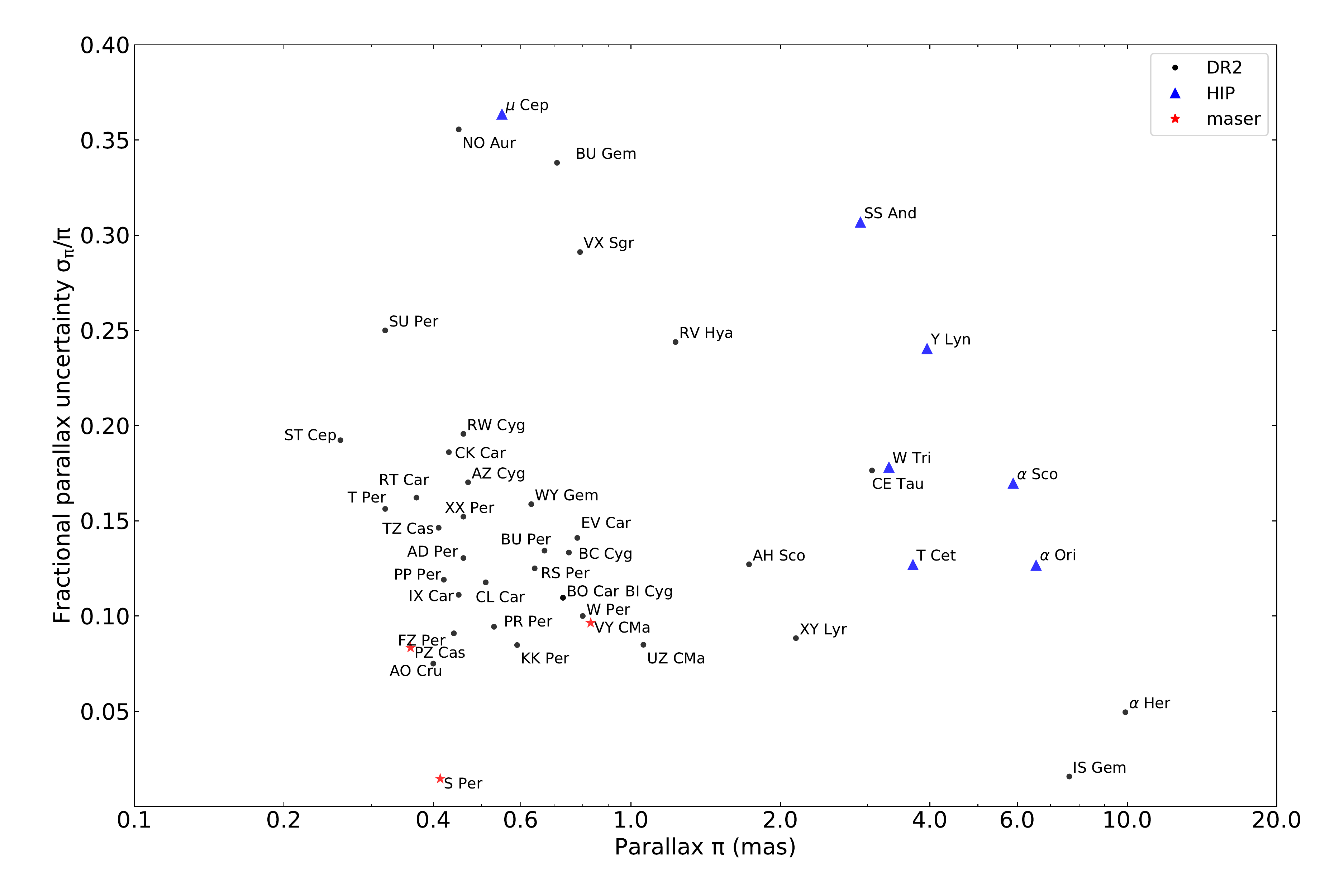}
\caption{Fractional parallax uncertainty measurements taken by Gaia DR2, Hipparcos and masers (see Section\:{\ref{sec:DR2}} for details). W~Ind and W~Cep with $\sigma_\pi$/$\pi$\,=~$0.57$ and $1.0$, respectively, have not been shown. BO~Car and BI~Cyg, with the same parallax and uncertainties values ($\pi$\,=~$0.73$ mas and $0.08$ mas), are indicated by one mark.}
\label{fig:fractplx}
\end{figure*}

Some targets have parallaxes from multiple sources, in which case we used the measurement with the smallest uncertainty. These are shown in Fig.\,{\ref{fig:fractplx}}. We took $37$ parallaxes from Gaia DR2, seven from Hipparcos \citep{Leeuwen+2007} and one from Gaia DR1 \citep[DR1;][]{Gaia+DR1}. Two objects, W~Ind and W~Cep, have large fractional uncertainties of $0.57$ (from DR1) and 1.0 (from DR2), respectively, and these stars have not been shown in Fig.\,{\ref{fig:fractplx}}.

We calculated distances by inverting the parallaxes. Because this can be a biased distance estimator (\citealt{LutzKelker+1973}, \citealt{BailerJones+2018, DR2+astrometric+2018, DR2+parallaxes+2018}), we compared these with distances from the \citet{BailerJones+2018} catalogue. Note that the \citet{BailerJones+2018} included a global parallax zero-point of $-0.029$\,mas (\citealt{DR2+astrometric+2018}). Once this was taken into account, we found excellent agreement, which confirms that for small values of fractional parallax (<$0.2$), the \citet{BailerJones+2018} posteriors are approximately Gaussian with a mode close to the inverse parallax \citep{BailerJones+2015,BailerJones+2018}. Since other zero-point offset values have been suggested for the DR2 parallaxes (\citealt{DR2+astrometric+2018, Riess+2018, Stello_offset+2018, Stassun+2018, Khan+2019}), we proceeded with the inverse parallax distances, without any correction. However, we consider the impact of the zero-point offset on the calculated absolute magnitudes of the Galactic RSGs and their P--L relation in Sec.\:{\ref{sec:RSGsinMW}}.

The majority of the sample (75\%) have fractional parallax uncertainties below $0.20$ and the so-called renormalized unit weight error (RUWE) below $1.4$ (threshold from \citealt{DR2+astrometric+2018}), which is what the Gaia team recommends when filtering on the unit weight error described in appendix C of \citealt{DR2+astrometric+2018}. However, we need to treat the DR2 uncertainties with caution because the associated astrometric measurements (excess noise, excess noise significance and goodness-of-fit) for these stars indicated low--quality fits. This may result from large--scale convective motions, which generate surface brightness and colour asymmetries, causing a shift of the photocentre that Gaia measures (\citealt{Chiavassa_AGB+2018}). Saturation could be another reason for large uncertainties, since $33$ Galactic RSGs have G\,<\,$7$\,mag, with $14$ stars brighter than $6$\,mag.

Table\:{\ref{table:MW Table}} shows the DR2 parallaxes and uncertainties used in the analysis. 
We calculated (and showed in Table\:{\ref{table:MW Table}}) the upper limit on the $M_K$ of W~Cep by assuming 0.1\,mas as an upper limit on the parallax. Finally, there are three objects, S~Per \citep{Asaki+2010}, VY~CMa \citep{Zhang+2012} and PZ~Cas \citep{Kusuno+2013}, that have accurate trigonometric parallaxes determined from measurements of the H\textsubscript{2}O (S~Per, PZ~Cas) and SiO (VY~CMa) masers.

%Unlike for Gaia DR1 , the parallax uncertainties have not been calibrated externally, i.e. they are known, as an ensemble, to be underestimated by $\sim$$8$-$12$\% for faint sources (G $\gtrsim$16 mag) outside the Galactic plane and by up to $\sim$30\% for bright stars (G $\lesssim$12 mag) (\citealt{DR2+parallaxes+2018}).

\subsection{Apparent $K$-band magnitudes}
\label{sec:ApMAg} % used for referring to this section from elsewhere

Photometry of red supergiants can be significantly affected by interstellar and/or circumstellar dust (\citealt{Josselin+2000}, \citealt{Massey+2005}). 
$V$ band data typically shows a larger spread in the observed magnitudes than the near--infrared (NIR) photometry, where the bolometric and extinction corrections are smaller (\citealt{Cardeli+1989}), and where changes in the observed variability due to pulsation are smaller (\citealt{Josselin+2000}; \citealt{Kiss++2006}; \citealt{Massey+2009}). In addition, fluxes of these objects peak in the NIR, and for these reasons, we followed previous studies of the P--L relation of RSGs in using $K$\,-band magnitudes.

To collect $K$-band apparent magnitudes, we followed the process described in \citet{Tabur+2009} and searched the available NIR catalogues: Catalogue of Infrared Observations (CIO, \citealt{Gezari+1999}) and Diffuse Infrared Background Experiment archive (DIRBE Catalogue of Stellar Photometry in Johnson{\textquotesingle}s,  \citealt{JohnsonsCatalogue}). We used the Gezari $K$-band magnitudes for $43$ stars and the Two Micron All Sky Survey (2MASS) catalogue (\citealt{Cutri+2003}) for three objects (CK~Car, AO~Cru and PP~Per). EV~Car and UZ~CMa were sourced in Catalogue of Stellar Photometry in Johnson's $11$-color system and DIRBE data, respectively. We combined the CIO observed magnitudes (at a wavelength of $2.2$\;{${\pm}$}\,$0.05$\;${\mu}$m) to calculate a median $K$ magnitude for each star, weighted equally because the catalogue did not provide uncertainties of the measurements. For the majority of our sample, $K$ magnitude uncertainties were not published. We calculated absolute magnitudes using the relation ${M = m + 5 + 5}$\,log\,$\pi$, where $m$ is the apparent magnitude, and $\pi$ is the parallax in arcseconds.
The values are presented in Table\:{\ref{table:MW Table}}, with further discussion in Section\:{\ref{sec:RSGsinMW}}.

\section{RED SUPERGIANTS IN THE LMC}
\label{sec:LMCsample}

\label{sec:DASCH} % used for referring to this section from elsewhere

The Digital Access to Sky Century @ Harvard \citep[DASCH;][]{DASCH+2012,DASCH+2013} program is digitising the Harvard Astronomical Plate Collection, which consists of tens of thousands of photographic plates spanning most of the $20$th century. These include the plates in the Small Magellanic Cloud that Henrietta Swan Leavitt used to determine the P--L relation in Cepheid variables (\citealt{Leavitt+Pickering+1912}). Our work uses the Large Magellanic Cloud plates, which have been recently digitised. Most stars we used have observation spanning approximately $10\,000$--$20\,000$ days, or $30$--$55$\,yr.

We obtained our list of $206$ supergiants primarily from \citet{YangJiang+2011} ($190$ objects), with the other objects taken from \citet{Massey+LMC+2003}\:(No.$205$, $208$, $209$, $210$, $216$ and $217$), \citet{Boyer+2011}\:(No.$191$, $195$, $196$, $199$ and $200$), \citet{CatchpoleFeast+1981}\:(No.$227$), \citet{Kastner+2008}\:(No.$219$, $220$ and $225$) and \citet{Levesque+2006}\:(No.$223$). We queried the DASCH online database \footnote{\url{http://dasch.rc.fas.harvard.edu/}} with a search radius of $5$\,arcsec and determined the correct star by looking at the magnitude and position. We extracted time and brightness data (uncertainties were not used in this study) and processed them using the same method as the Galactic sample (see Sec.\:{\ref{sec:period}}), except a bin size of $10$\,d was used, since the available time span of the DASCH observations was shorter than the AAVSO data. We trimmed each lightcurve so it only contained measurements on the interval $2\,420\,000$$-$$2\,440\,000$\,(JD), where the majority of useful observations occurred.

We completed Fourier analysis using the same methodology and software that were used for the Galactic RSGs. We searched for frequencies between $0.000125$ and $0.01\,$d\textsuperscript{-1}, corresponding to periods of $100$--$8000$\,d.

\begin{figure*}
\centering

\begin{subfigure}[b]{0.7\textwidth}
   \includegraphics[width=\linewidth]{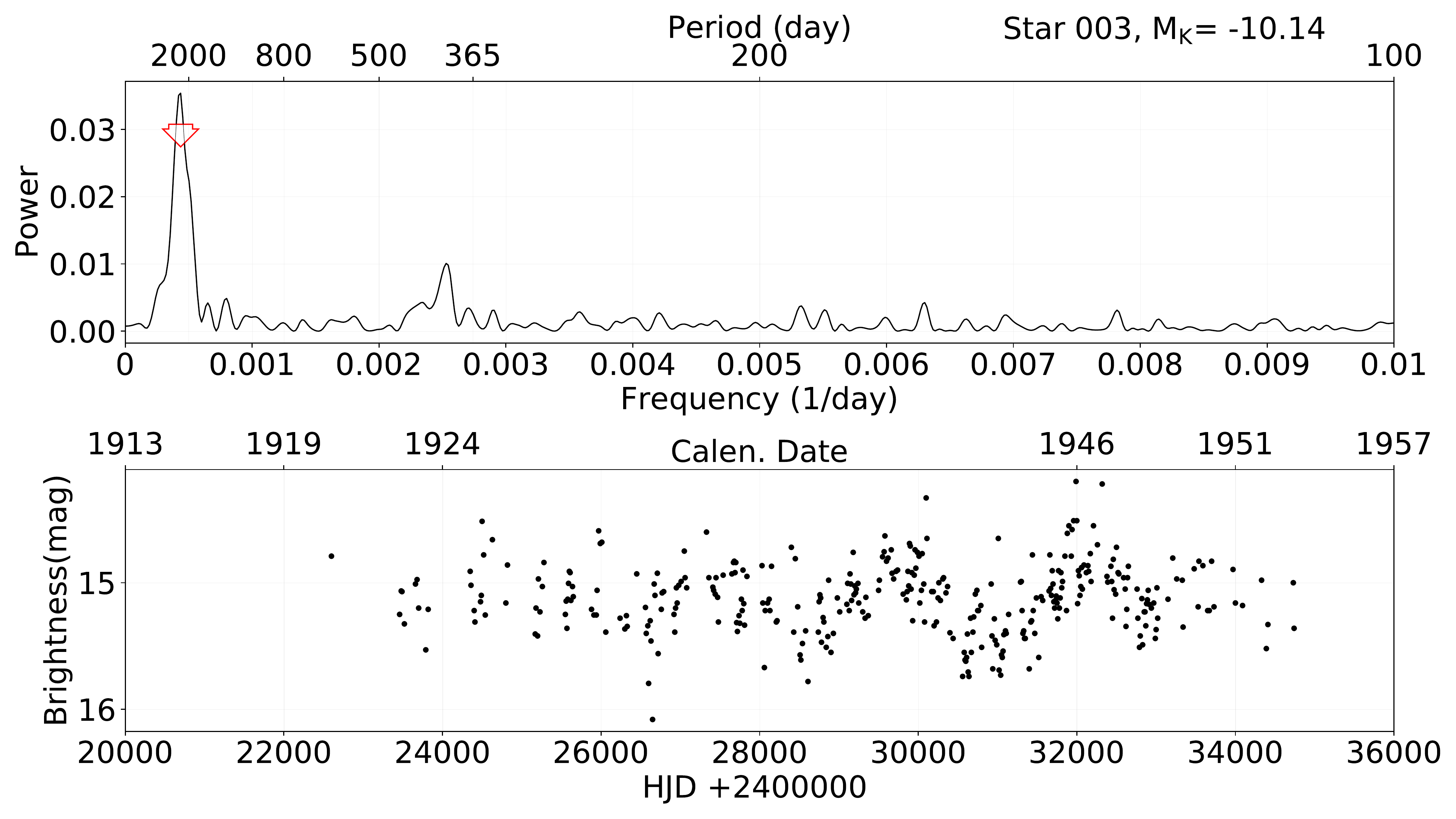}

\end{subfigure}\vspace{8mm}

\begin{subfigure}[b]{0.7\textwidth}
   \includegraphics[width=\linewidth]{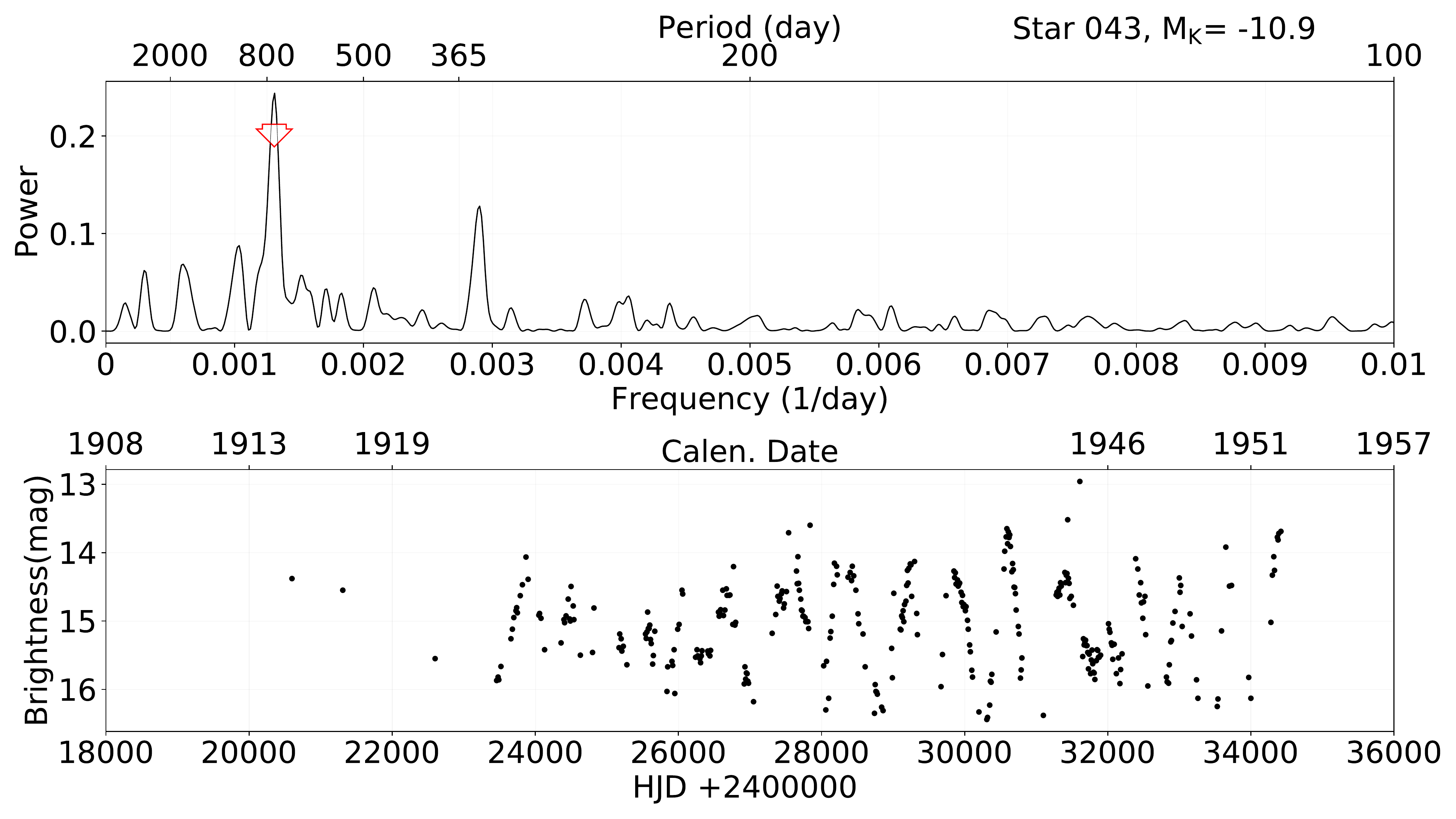}
   
  \end{subfigure}\vspace{8mm}

\begin{subfigure}[b]{0.7\textwidth}
   \includegraphics[width=\linewidth]{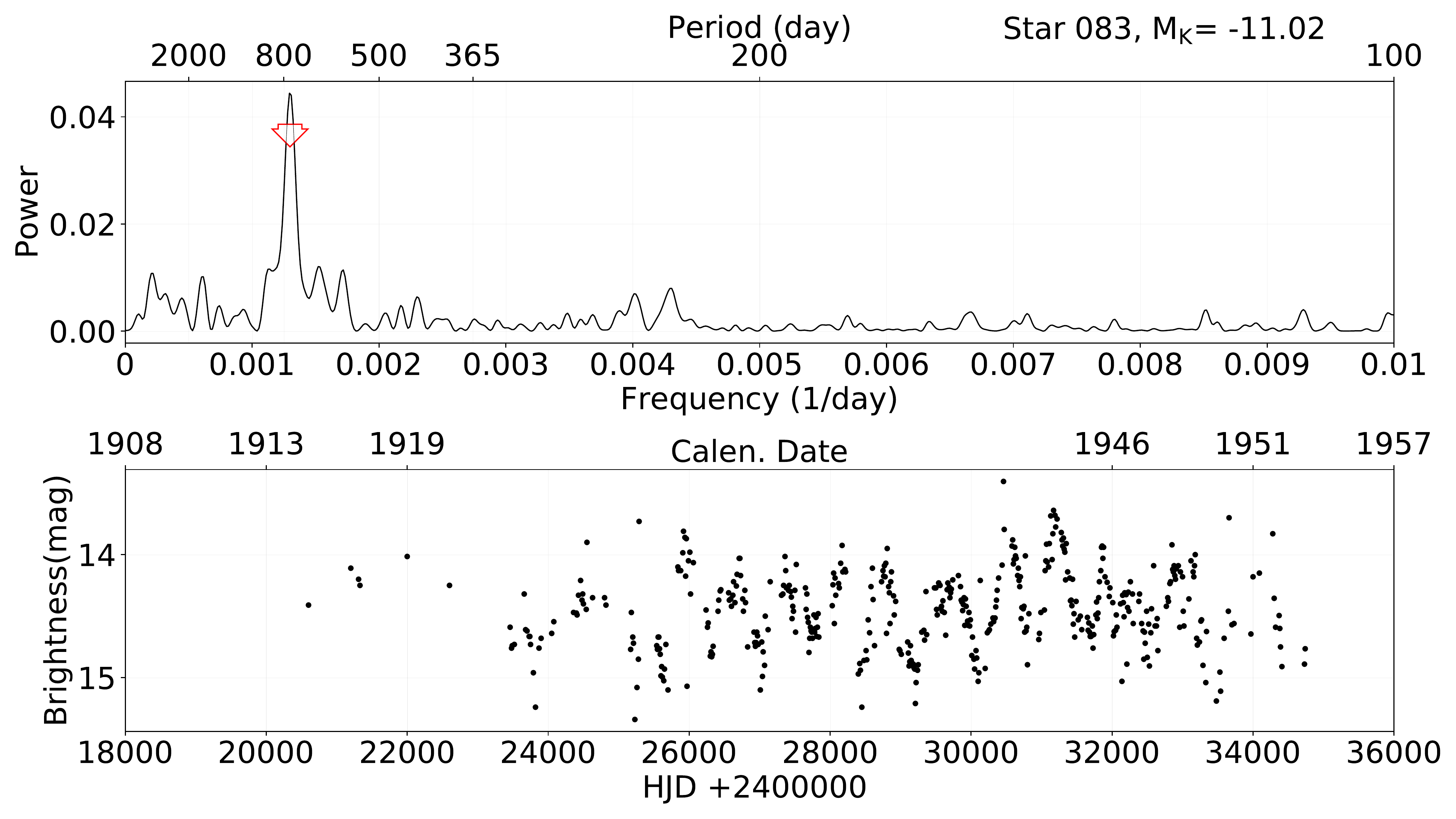}
 
  \end{subfigure}

\caption[]{Sample DASCH ($10$\,-d bins) lightcurves with associated power density spectra of three RSGs in the LMC, 003, 043 and 083. Arrows indicate measured pulsational periods (All lightcurves are available in the supplementary material).}
\label{fig:LMCPS}
\end{figure*}

Out of the $206$ objects we selected $170$ lightcurves for further analysis; the other $36$ were either not available in DASCH or had poor-quality or sparse lightcurves. We inspected the power spectra and lightcurves of these $170$ objects and found periods in $142$ ($83$\%). The rest had power spectra dominated by 1/$f$ noise with no clear periodicity. In Fig.\:{\ref{fig:LMCPS}} we show three examples (stars No.003, 043, 083) of the lightcurves and power spectra with identified periods.

\begin{figure}
\includegraphics[width=\linewidth]{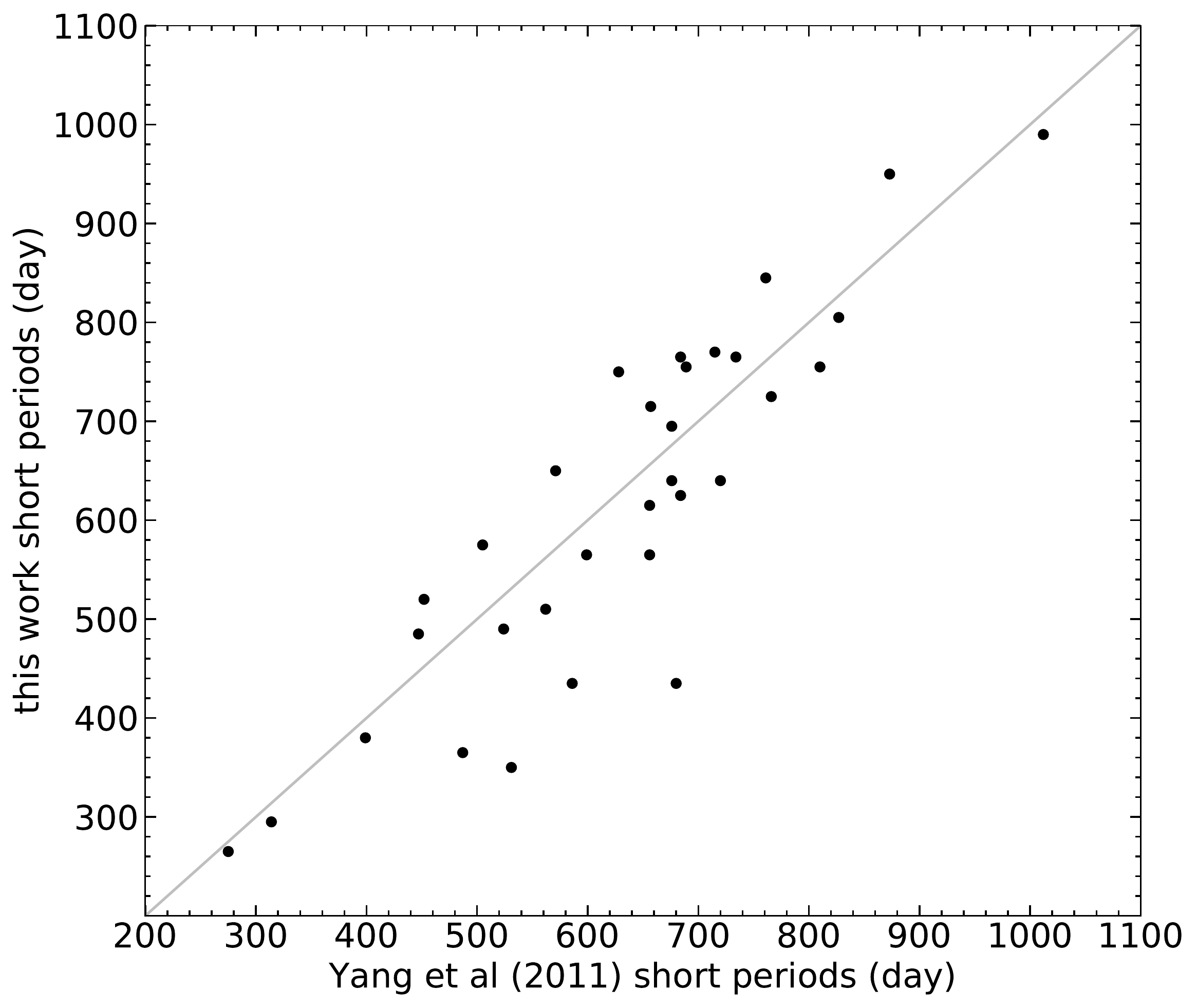}
\caption{Comparison of shorter periods of $32$ RSGs in the LMC, from this study with periods published by Yang et al. (2011)\;. The grey line shows equality.}
\label{fig:YangvsChatys}
\end{figure}

In Figure\,{\ref{fig:YangvsChatys}} we compare the DASCH shorter periods of $32$ LMC RSGs, with values from \citet{YangJiang+2011}. For the majority of objects ($81$\%), these agree  to within $15$\%. Disagreement can result from the stochastic nature of periodicity if observations are taken at different times. Because DASCH data have longer timespans, we used our periods in preference to the literature values. We note that we do not show LSPs, published by \citet{YangJiang+2011} because they were based on only $3000$\,d of data (mostly ASAS), which we consider too short for a reliable detection of LSPs.

Table\:{\ref{table:LMC Table}} shows the LMC sample with stars numbered as per \citet{YangJiang+2011}. Columns $2$ and $3$ present coordinates of the stars, followed by identified periods, the associated amplitudes, $K$\,magnitudes, their uncertainties and the absolute $K$\,magnitudes.
We list all $206$ objects that form our LMC sample. Stars without periods can be classified into four groups -- (i) those without enough data to perform a reasonable Fourier transform, (ii) those with high noise in the Fourier transform (making it unreasonable to determine a period), (iii) those with unreliable periods $\gtrsim$ $8\,000$\,d, and (iv) six objects for which the DASCH database did not provide a time series.

We took $K$-band magnitudes of the LMC stars from the 2MASS catalogue (\citealt{Cutri+2003}). Most stars have a magnitude uncertainty of less than $0.03$\,mag, which is shown in Table\:{\ref{table:LMC Table}}.
We adopted a distance modulus of ${\mu}($LMC)\,= $18.476$\,mag (\citealt{Pietrzynski_LMC+2019}).

\begin{table*}
\caption[]{Coordinates, $K$-band magnitude, period and amplitude of our sample of RSGs in the LMC. 
Sources as per Section\:{\ref{sec:LMCsample}}.}
\label{table:LMC Table}
\begin{tabular}{@{\extracolsep{\fill}}l l l c c c c c r c c c r}
\hline
\hline
Ref. & $\alpha_{J2000}$ & $\delta_{J2000}$ & P1 &amp1& P2&amp2&P3 &amp3&$K$&$\sigma_{K}$&$M_K$\\
& & &  (days) & (mag)& (days) & (mag)& (days) & (mag) & (mag) & (mag) & (mag)\\
\hline
001 & $72.3436$ & $-69.4096$ & $610$ & $0.20$ & $1950$ & $0.16$ & $--$ & --- & $7.76$ & $0.04$ & $-10.74$ \\

003 & $72.7445$ & $-69.2341$ & $2290$ & $0.19$ & $--$ & --- & $--$ & --- & $8.36$ & $0.02$ & $-10.14$ \\

004 & $72.8792$ & $-69.2478$ & $490$ & $0.12$ & $674$ & $0.10$ & $2681$ & $0.15$ & $8.24$ & $0.02$ & $-10.26$ \\

005 & $72.9470$ & $-69.3235$ & $365$ & $0.15$ & $--$ & --- & $--$ & --- & $8.74$ & $0.02$ & $-9.76$ \\

006 & $73.3116$ & $-69.2050$ & $--$ & --- & $--$ & --- & $--$ & --- & $7.75$ & $0.02$ & $-10.75$ \\

007 & $73.3269$ & $-69.2842$ & $5000$ & $0.06$ & $--$ & --- & $--$ & --- & $8.36$ & $0.02$ & $-10.14$ \\

008 & $73.3788$ & $-69.2971$ & $--$ & --- & $--$ & --- & $--$ & --- & $8.06$ & $0.02$ & $-10.44$ \\

009 & $73.6536$ & $-69.3395$ & $4340$ & $0.07$ & $--$ & --- & $--$ & --- & $7.61$ & $0.03$ & $-10.89$ \\

010 & $73.6606$ & $-69.1881$ & $755$ & $0.16$ & $2350$ & $0.17$ & $--$ & --- & $7.20$ & $0.03$ & $-11.30$ \\

011 & $73.6642$ & $-69.0768$ & $3970$ & $0.56$ & $--$ & --- & $--$ & --- & $8.66$ & $0.02$ & $-9.84$ \\

012 & $73.7070$ & $-69.5007$ & $1560$ & $0.08$ & $--$ & --- & $--$ & --- & $8.43$ & $0.02$ & $-10.07$ \\

014 & $73.8169$ & $-69.3200$ & $435$ & $0.08$ & $1500$ & $0.07$ & $--$ & --- & $7.37$ & $0.03$ & $-11.13$ \\

016 & $73.8750$ & $-69.4863$ & $565$ & $0.18$ & $--$ & --- & $--$ & --- & $7.66$ & $0.02$ & $-10.84$ \\

017 & $73.8836$ & $-66.8439$ & $715$ & $0.17$ & $3070$ & $0.16$ & $--$ & --- & $7.66$ & $0.02$ & $-10.84$ \\

019 & $73.9243$ & $-69.4401$ & $610$ & $0.27$ & $--$ & --- & $--$ & --- & $7.70$ & $0.02$ & $-10.80$ \\

020 & $73.9511$ & $-69.4018$ & $335$ & $0.08$ & $--$ & --- & $--$ & --- & $7.97$ & $0.02$ & $-10.53$ \\

021 & $74.0986$ & $-69.7031$ & $320$ & $0.06$ & $2510$ & $0.07$ & $--$ & --- & $8.45$ & $0.02$ & $-10.05$ \\

023 & $74.3814$ & $-70.1498$ & $610$ & $0.23$ & $--$ & --- & $--$ & --- & $7.97$ & $0.03$ & $-10.53$ \\

024 & $74.4305$ & $-70.1473$ & $950$ & $0.27$ & $--$ & --- & $--$ & --- & $7.32$ & $0.03$ & $-11.18$ \\

025 & $74.4358$ & $-69.5095$ & $--$ & --- & $--$ & --- & $--$ & --- & $8.56$ & $0.02$ & $-9.94$ \\

026 & $75.5397$ & $-70.4172$ & $--$ & --- & $--$ & --- & $--$ & --- & $8.32$ & $0.02$ & $-10.18$ \\

027 & $75.8137$ & $-70.2949$ & $2375$ & $0.06$ & $--$ & --- & $--$ & --- & $8.72$ & $0.02$ & $-9.78$ \\

028 & $76.0210$ & $-70.3796$ & $4100$ & $0.13$ & $--$ & --- & $--$ & --- & $8.11$ & $0.03$ & $-10.39$ \\

029 & $76.0409$ & $-70.2049$ & $3413$ & $0.06$ & $322$ & $0.05$ & $--$ & --- & $8.39$ & $0.03$ & $-10.11$ \\

030 & $76.0588$ & $-67.2707$ & $990$ & $0.25$ & $--$ & --- & $--$ & --- & $6.78$ & $0.02$ & $-11.72$ \\

031 & $76.1741$ & $-70.7104$ & $2950$ & $0.17$ & $--$ & --- & $--$ & --- & $8.03$ & $0.03$ & $-10.47$ \\

032 & $76.2259$ & $-70.5552$ & $556$ & $0.15$ & $--$ & --- & $--$ & --- & $8.40$ & $0.02$ & $-10.10$ \\

033 & $76.2916$ & $-70.6677$ & $405$ & $0.13$ & $--$ & --- & $--$ & --- & $8.38$ & $0.02$ & $-10.12$ \\

034 & $76.3897$ & $-70.5630$ & $695$ & $0.24$ & $--$ & --- & $--$ & --- & $7.64$ & $0.03$ & $-10.86$ \\

035 & $76.4863$ & $-70.5900$ & $2760$ & $0.19$ & $--$ & --- & $--$ & --- & $8.11$ & $0.03$ & $-10.39$ \\

037 & $76.4981$ & $-70.8032$ & $3675$ & $0.27$ & $--$ & --- & $--$ & --- & $8.32$ & $0.02$ & $-10.18$ \\

038 & $76.6517$ & $-70.5441$ & $2667$ & $0.14$ & $--$ & --- & $--$ & --- & $8.75$ & $0.02$ & $-9.75$ \\

039 & $76.7737$ & $-70.5456$ & $640$ & $0.12$ & $--$ & --- & $--$ & --- & $7.04$ & $0.03$ & $-11.46$ \\

040 & $76.8857$ & $-70.6512$ & $520$ & $0.10$ & $--$ & --- & $--$ & --- & $8.02$ & $0.02$ & $-10.48$ \\

042 & $77.4317$ & $-65.3665$ & $615$ & $0.17$ & $--$ & --- & $--$ & --- & $7.69$ & $0.03$ & $-10.81$ \\

043 & $78.1932$ & $-67.3272$ & $765$ & $0.49$ & $--$ & --- & $--$ & --- & $7.59$ & $0.03$ & $-10.91$ \\

044 & $78.7072$ & $-67.4555$ & $805$ & $0.44$ & $--$ & --- & $--$ & --- & $7.42$ & $0.02$ & $-11.08$ \\

045 & $79.2874$ & $-69.5392$ & $575$ & $0.18$ & $265$ & $0.15$ & $--$ & --- & $7.82$ & $0.02$ & $-10.68$ \\

046 & $79.4848$ & $-69.6737$ & $835$ & $0.07$ & $--$ & --- & $--$ & --- & $8.57$ & $0.02$ & $-9.93$ \\

047 & $79.7636$ & $-69.6653$ & $3650$ & $0.33$ & $--$ & --- & $375$ & $0.53$ & $7.60$ & $0.03$ & $-10.90$ \\

048 & $79.9720$ & $-69.4593$ & $1750$ & $0.15$ & $--$ & --- & $--$ & --- & $8.22$ & $0.03$ & $-10.28$ \\

049 & $80.0984$ & $-69.5575$ & $520$ & $0.18$ & $--$ & --- & $--$ & --- & $7.98$ & $0.02$ & $-10.52$ \\

050 & $80.3665$ & $-69.5045$ & $1635$ & $0.13$ & $--$ & --- & $--$ & --- & $8.14$ & $0.02$ & $-10.36$ \\

051 & $80.6295$ & $-69.5681$ & $--$ & --- & $--$ & --- & $--$ & --- & $8.65$ & $0.02$ & $-9.85$ \\

052 & $80.7615$ & $-69.3436$ & $285$ & $0.06$ & $--$ & --- & $--$ & --- & $8.51$ & $0.02$ & $-9.99$ \\

053 & $80.8916$ & $-69.3186$ & $1250$ & $0.08$ & $--$ & --- & $380$ & $0.09$ & $8.90$ & $0.02$ & $-9.60$ \\

054 & $80.9317$ & $-65.6999$ & $675$ & $0.28$ & $--$ & --- & $--$ & --- & $7.74$ & $0.05$ & $-10.76$ \\

056 & $81.0804$ & $-69.6470$ & $4900$ & $0.12$ & $--$ & --- & $--$ & --- & $6.81$ & $0.02$ & $-11.69$ \\

057 & $81.4369$ & $-69.0802$ & $510$ & $0.13$ & $2470$ & $0.13$ & $--$ & --- & $7.99$ & $0.02$ & $-10.51$ \\

060 & $81.5671$ & $-66.1164$ & $510$ & $0.27$ & $--$ & --- & $--$ & --- & $8.03$ & $0.03$ & $-10.47$ \\

061 & $81.6141$ & $-69.1822$ & $655$ & $0.37$ & $3510$ & $0.23$ & $--$ & --- & $7.70$ & $0.02$ & $-10.80$ \\

062 & $81.6176$ & $-69.1327$ & $405$ & $0.11$ & $2650$ & $0.10$ & $--$ & --- & $8.48$ & $0.02$ & $-10.02$ \\

063 & $81.6450$ & $-68.8611$ & $3210$ & $0.19$ & $--$ & --- & $--$ & --- & $7.26$ & $0.03$ & $-11.24$ \\

065 & $81.6780$ & $-68.9536$ & $435$ & $0.14$ & $2250$ & $0.12$ & $--$ & --- & $8.55$ & $0.02$ & $-9.95$ \\

067 & $81.7929$ & $-69.2715$ & $3700$ & $0.26$ & $--$ & --- & $--$ & --- & $8.78$ & $0.02$ & $-9.72$ \\

070 & $81.8669$ & $-69.0100$ & $3670$ & $0.57$ & $--$ & --- & $--$ & --- & $8.32$ & $0.02$ & $-10.18$ \\

071 & $81.8737$ & $-67.2370$ & $1230$ & $0.15$ & $--$ & --- & $--$ & --- & $7.97$ & $0.03$ & $-10.53$ \\

072 & $81.8931$ & $-66.8917$ & $650$ & $0.23$ & $--$ & --- & $--$ & --- & $7.84$ & $0.03$ & $-10.66$ \\

074 & $81.9479$ & $-69.2224$ & $3735$ & $0.50$ & $--$ & --- & $--$ & --- & $7.60$ & $0.03$ & $-10.90$ \\

075 & $81.9630$ & $-67.3011$ & $--$ & --- & $--$ & --- & $--$ & --- & $8.60$ & $0.02$ & $-9.90$ \\

076 & $81.9631$ & $-69.1794$ & $1250$ & $0.09$ & $--$ & --- & $--$ & --- & $8.29$ & $0.03$ & $-10.21$ \\

077 & $82.0249$ & $-69.1204$ & $1270$ & $0.13$ & $--$ & --- & $--$ & --- & $8.15$ & $0.03$ & $-10.35$ \\

\hline

\end{tabular}
\end{table*}

\begin{table*}
\contcaption{}

\label{tab:continued}
\begin{tabular}{@{\extracolsep{\fill}}l l l c c c c c r c c c r}
\hline
\hline
Ref. & $\alpha_{J2000}$ & $\delta_{J2000}$ & P1 &amp1& P2&amp2&P3 &amp3&$K$&$\sigma_{K}$&$M_K$\\
& & &  (days) & (mag)& (days) & (mag)& (days) & (mag) & (mag) & (mag) & (mag)\\
\hline
079 & $82.0642$ & $-66.9813$ & $510$ & $0.18$ & $4100$ & $0.16$ & $--$ & --- & $8.09$ & $0.02$ & $-10.41$ \\

080 & $82.0662$ & $-69.2003$ & $--$ & --- & $--$ & --- & $--$ & --- & $8.51$ & $0.02$ & $-9.99$ \\

081 & $82.0775$ & $-69.1264$ & $1665$ & $0.34$ & $--$ & --- & $--$ & --- & $8.31$ & $0.03$ & $-10.19$ \\

082 & $82.1163$ & $-69.2159$ & $3735$ & $0.20$ & $--$ & --- & $370$ & $0.21$ & $8.38$ & $0.03$ & $-10.12$ \\

083 & $82.1202$ & $-68.1189$ & $770$ & $0.21$ & $--$ & --- & $--$ & --- & $7.48$ & $0.02$ & $-11.02$ \\

084 & $82.1264$ & $-69.0123$ & $--$ & --- & $--$ & --- & $--$ & --- & $8.50$ & $0.02$ & $-10.00$ \\

085 & $82.1314$ & $-69.0920$ & $480$ & $0.16$ & $--$ & --- & $--$ & --- & $8.05$ & $0.03$ & $-10.45$ \\

087 & $82.1799$ & $-67.3079$ & $2570$ & $0.10$ & $--$ & --- & $--$ & --- & $8.58$ & $0.02$ & $-9.92$ \\

091 & $82.2532$ & $-68.7760$ & $--$ & --- & $--$ & --- & $--$ & --- & $8.44$ & $0.02$ & $-10.06$ \\

092 & $82.2645$ & $-69.1128$ & $640$ & $0.26$ & $--$ & --- & $--$ & --- & $7.90$ & $0.02$ & $-10.60$ \\

093 & $82.2729$ & $-67.3049$ & $2275$ & $0.04$ & $--$ & --- & $--$ & --- & $8.57$ & $0.02$ & $-9.93$ \\

094 & $82.2850$ & $-69.2051$ & $3390$ & $0.24$ & $--$ & --- & $370$ & $0.27$ & $8.35$ & $0.04$ & $-10.15$ \\

097 & $82.3650$ & $-69.1473$ & $765$ & $0.31$ & $--$ & --- & $--$ & --- & $7.30$ & $0.02$ & $-11.20$ \\

098 & $82.3934$ & $-66.9245$ & $500$ & $0.40$ & $--$ & --- & $--$ & --- & $8.73$ & $0.02$ & $-9.77$ \\

099 & $82.4258$ & $-68.9548$ & $845$ & $0.15$ & $--$ & --- & $--$ & --- & $6.89$ & $0.03$ & $-11.61$ \\

100 & $82.4332$ & $-69.0972$ & $520$ & $0.18$ & $3775$ & $0.16$ & $--$ & --- & $7.88$ & $0.02$ & $-10.62$ \\

101 & $82.4782$ & $-69.0710$ & $295$ & $0.06$ & $--$ & --- & $--$ & --- & $8.41$ & $0.02$ & $-10.09$ \\

102 & $82.4789$ & $-67.3102$ & $750$ & $0.09$ & $--$ & --- & $365$ & $0.10$ & $7.79$ & $0.02$ & $-10.71$ \\

103 & $82.5095$ & $-67.0459$ & $575$ & $0.12$ & $--$ & --- & $--$ & --- & $7.97$ & $0.02$ & $-10.53$ \\

104 & $82.5191$ & $-68.7913$ & $3445$ & $0.06$ & $--$ & --- & $--$ & --- & $8.77$ & $0.02$ & $-9.73$ \\

105 & $82.5206$ & $-69.0666$ & $810$ & $0.12$ & $1900$ & $0.11$ & $--$ & --- & $8.81$ & $0.02$ & $-9.69$ \\

106 & $82.5399$ & $-69.1844$ & $3935$ & $0.09$ & $--$ & --- & $--$ & --- & $8.64$ & $0.02$ & $-9.86$ \\

107 & $82.5873$ & $-67.3348$ & $625$ & $0.14$ & $2665$ & $0.16$ & $--$ & --- & $7.45$ & $0.03$ & $-11.05$ \\

108 & $82.5921$ & $-67.1088$ & $3570$ & $0.29$ & $--$ & --- & $--$ & --- & $8.60$ & $0.02$ & $-9.90$ \\

109 & $82.6095$ & $-69.5068$ & $2235$ & $0.23$ & $--$ & --- & $--$ & --- & $8.48$ & $0.02$ & $-10.02$ \\

111 & $82.6481$ & $-68.9898$ & $365$ & $0.19$ & $4330$ & $0.25$ & $670$ & --- & $7.55$ & $0.02$ & $-10.95$ \\

112 & $82.6482$ & $-67.2012$ & $1160$ & $0.07$ & $--$ & --- & $--$ & --- & $8.85$ & $0.02$ & $-9.65$ \\

113 & $82.6727$ & $-69.2594$ & $4265$ & $0.33$ & $--$ & --- & $--$ & --- & $7.59$ & $0.02$ & $-10.91$ \\

114 & $82.6749$ & $-69.0898$ & $2155$ & $0.13$ & $--$ & --- & $--$ & --- & $8.76$ & $0.02$ & $-9.74$ \\

115 & $82.6882$ & $-67.1332$ & $190$ & $0.10$ & $2770$ & $0.12$ & $--$ & --- & $8.43$ & $0.02$ & $-10.07$ \\

116 & $82.7179$ & $-67.2929$ & $200$ & $0.10$ & $2610$ & $0.11$ & $--$ & --- & $8.83$ & $0.02$ & $-9.67$ \\

118 & $82.7550$ & $-69.1831$ & $365$ & $0.22$ & $--$ & --- & $--$ & --- & $8.33$ & $0.03$ & $-10.17$ \\

119 & $82.7643$ & $-69.0945$ & $--$ & --- & $--$ & --- & $--$ & --- & $8.58$ & $0.03$ & $-9.92$ \\

120 & $82.7674$ & $-69.3175$ & $652$ & $0.26$ & $--$ & --- & $--$ & --- & $7.63$ & $0.03$ & $-10.87$ \\

121 & $82.7886$ & $-67.4319$ & $565$ & $0.19$ & $3175$ & $0.19$ & $--$ & --- & $7.63$ & $0.02$ & $-10.87$ \\

122 & $82.8144$ & $-69.0664$ & $3255$ & $0.09$ & $--$ & --- & $--$ & --- & $8.22$ & $0.02$ & $-10.28$ \\

123 & $82.8269$ & $-69.1578$ & $--$ & --- & $--$ & --- & $--$ & --- & $8.63$ & $0.02$ & $-9.87$ \\

124 & $82.9034$ & $-66.5021$ & $725$ & $0.30$ & $--$ & --- & $--$ & --- & $7.37$ & $0.02$ & $-11.13$ \\

125 & $82.9475$ & $-67.3842$ & $--$ & --- & $--$ & --- & $--$ & --- & $8.59$ & $0.02$ & $-9.91$ \\

126 & $83.0367$ & $-67.1885$ & $--$ & --- & $--$ & --- & $--$ & --- & $8.69$ & $0.03$ & $-9.81$ \\

128 & $83.1143$ & $-69.2813$ & $465$ & $0.14$ & $--$ & --- & $--$ & --- & $7.96$ & $0.02$ & $-10.54$ \\

129 & $83.1306$ & $-69.3404$ & $965$ & $0.06$ & $--$ & --- & $365$ & $0.07$ & $8.63$ & $0.02$ & $-9.87$ \\

130 & $83.1471$ & $-69.1310$ & $--$ & --- & $--$ & --- & $--$ & --- & $8.26$ & $0.02$ & $-10.24$ \\

131 & $83.2093$ & $-67.4625$ & $440$ & $0.12$ & $--$ & --- & $--$ & --- & $8.05$ & $0.03$ & $-10.45$ \\

132 & $83.2817$ & $-66.8016$ & $485$ & $0.22$ & $2350$ & $0.22$ & $--$ & --- & $8.61$ & $0.02$ & $-9.89$ \\

134 & $83.3617$ & $-67.0704$ & $350$ & $0.07$ & $1669$ & $0.07$ & $--$ & --- & $7.82$ & $0.03$ & $-10.68$ \\

135 & $83.3733$ & $-67.5271$ & $--$ & --- & $--$ & --- & $--$ & --- & $8.82$ & $0.02$ & $-9.68$ \\

136 & $83.4356$ & $-67.4047$ & $380$ & $0.08$ & $2350$ & $0.08$ & $--$ & --- & $8.49$ & $0.02$ & $-10.01$ \\

137 & $83.4674$ & $-69.1871$ & $645$ & $0.20$ & $--$ & --- & $--$ & --- & $7.90$ & $0.02$ & $-10.60$ \\

138 & $83.5586$ & $-68.9789$ & $--$ & --- & $--$ & --- & $--$ & --- & $7.96$ & $0.02$ & $-10.54$ \\

139 & $83.5812$ & $-68.9935$ & $--$ & --- & $--$ & --- & $--$ & --- & $8.53$ & $0.02$ & $-9.97$ \\

140 & $83.5893$ & $-69.3667$ & $295$ & $0.10$ & $1550$ & $0.09$ & $--$ & --- & $8.89$ & $0.02$ & $-9.61$ \\

141 & $83.6407$ & $-69.2507$ & $2665$ & $0.18$ & $--$ & --- & $--$ & --- & $8.28$ & $0.02$ & $-10.22$ \\

142 & $83.6959$ & $-69.4835$ & $--$ & --- & $--$ & --- & $--$ & --- & $9.02$ & $0.03$ & $-9.48$ \\

143 & $83.8088$ & $-67.7322$ & $710$ & $0.35$ & $--$ & --- & $--$ & --- & $8.02$ & $0.02$ & $-10.48$ \\

144 & $83.8288$ & $-67.0388$ & $415$ & $0.19$ & $575$ & $0.16$ & $2550$ & $0.16$ & $8.34$ & $0.03$ & $-10.16$ \\

145 & $83.8522$ & $-69.0676$ & $310$ & $0.12$ & $850$ & $0.15$ & $1950$ & $0.14$ & $8.23$ & $0.03$ & $-10.27$ \\

146 & $83.8680$ & $-66.9340$ & $755$ & $0.28$ & $--$ & --- & $--$ & --- & $7.26$ & $0.03$ & $-11.24$ \\

147 & $83.8867$ & $-69.0720$ & $--$ & --- & $--$ & --- & $--$ & --- & $8.20$ & $0.02$ & $-10.30$ \\

148 & $83.9326$ & $-68.8558$ & $2620$ & $0.13$ & $--$ & --- & $--$ & --- & $8.04$ & $0.02$ & $-10.46$ \\

149 & $83.9665$ & $-69.3748$ & $1700$ & $0.08$ & $--$ & --- & $--$ & --- & $8.45$ & $0.02$ & $-10.05$ \\

151 & $84.0266$ & $-68.9447$ & $--$ & --- & $--$ & --- & $--$ & --- & $8.44$ & $0.02$ & $-10.06$ \\

\hline

\end{tabular}
\end{table*}

\begin{table*}
\contcaption{.}

\label{tab:continued}
\begin{tabular}{@{\extracolsep{\fill}}l l l c c c c c r c c c r}
\hline
\hline
Ref. & $\alpha_{J2000}$ & $\delta_{J2000}$ & P1 &amp1& P2&amp2&P3 &amp3&$K$&$\sigma_{K}$&$M_K$\\
& & &  (days) & (mag)& (days) & (mag)& (days) & (mag) & (mag) & (mag) & (mag)\\
\hline
154 & $84.1061$ & $-66.9273$ & $750$ & $0.23$ & $--$ & --- & $--$ & --- & $7.50$ & $0.03$ & $-11.00$ \\

155 & $84.1115$ & $-69.3976$ & $465$ & $0.19$ & $--$ & --- & $--$ & --- & $7.81$ & $0.02$ & $-10.69$ \\

156 & $84.1691$ & $-69.3879$ & $--$ & --- & $--$ & --- & $--$ & --- & $8.81$ & $0.02$ & $-9.69$ \\

158 & $84.3599$ & $-68.7945$ & $1800$ & $0.10$ & $--$ & --- & $--$ & --- & $8.23$ & $0.02$ & $-10.27$ \\

159 & $84.3777$ & $-69.0425$ & $3560$ & $0.10$ & $--$ & --- & $--$ & --- & $8.72$ & $0.02$ & $-9.78$ \\

160 & $84.4037$ & $-69.4899$ & $--$ & --- & $--$ & --- & $--$ & --- & $8.21$ & $0.02$ & $-10.29$ \\

162 & $84.4379$ & $-69.3468$ & $--$ & --- & $--$ & --- & $--$ & --- & $7.72$ & $0.03$ & $-10.78$ \\

163 & $84.4945$ & $-69.2400$ & $--$ & --- & $--$ & --- & $--$ & --- & $8.38$ & $0.02$ & $-10.12$ \\

166 & $84.5755$ & $-69.2951$ & $1965$ & $0.13$ & $--$ & --- & $--$ & --- & $8.30$ & $0.02$ & $-10.20$ \\

167 & $84.6418$ & $-69.3422$ & $4560$ & $0.13$ & $--$ & --- & $--$ & --- & $8.51$ & $0.02$ & $-9.99$ \\

168 & $84.9426$ & $-69.3245$ & $--$ & --- & $--$ & --- & $--$ & --- & $8.47$ & $0.02$ & $-10.03$ \\

170 & $85.0320$ & $-69.3347$ & $2200$ & $0.14$ & $--$ & --- & $--$ & --- & $8.29$ & $0.03$ & $-10.21$ \\

172 & $85.1022$ & $-69.3548$ & $--$ & --- & $--$ & --- & $--$ & --- & $7.85$ & $0.03$ & $-10.65$ \\

173 & $85.1058$ & $-69.2584$ & $3215$ & $0.20$ & $--$ & --- & $--$ & --- & $8.78$ & $0.02$ & $-9.72$ \\

174 & $85.1541$ & $-69.4390$ & $--$ & --- & $--$ & --- & $--$ & --- & $8.32$ & $0.02$ & $-10.18$ \\

175 & $85.1826$ & $-69.3662$ & $1300$ & $0.16$ & $--$ & --- & $--$ & --- & $7.44$ & $0.02$ & $-11.06$ \\

177 & $85.2308$ & $-69.3904$ & $2570$ & $0.13$ & $--$ & --- & $--$ & --- & $7.54$ & $0.02$ & $-10.96$ \\

178 & $85.2470$ & $-69.3101$ & $715$ & $0.12$ & $2515$ & $0.13$ & $--$ & --- & $7.49$ & $0.03$ & $-11.01$ \\

179 & $85.2712$ & $-69.0784$ & $3445$ & $0.15$ & $--$ & --- & $--$ & --- & $7.98$ & $0.02$ & $-10.52$ \\

180 & $85.2789$ & $-69.2874$ & $2315$ & $0.16$ & $--$ & --- & $--$ & --- & $7.77$ & $0.02$ & $-10.73$ \\

181 & $85.2948$ & $-69.6345$ & $--$ & --- & $--$ & --- & $--$ & --- & $7.63$ & $0.02$ & $-10.87$ \\

182 & $85.3408$ & $-69.5303$ & $435$ & $0.16$ & $2455$ & $0.24$ & $--$ & --- & $7.82$ & $0.03$ & $-10.68$ \\

183 & $85.3731$ & $-69.4544$ & $2100$ & $0.07$ & $--$ & --- & $--$ & --- & $8.45$ & $0.02$ & $-10.05$ \\

184 & $85.4309$ & $-69.4710$ & $2778$ & $0.07$ & $--$ & --- & $--$ & --- & $8.41$ & $0.02$ & $-10.09$ \\

185 & $85.4335$ & $-69.2008$ & $3215$ & $0.21$ & $--$ & --- & $--$ & --- & $8.40$ & $0.02$ & $-10.10$ \\

186 & $85.4590$ & $-69.3543$ & $265$ & $0.07$ & $1530$ & $0.06$ & $--$ & --- & $8.56$ & $0.02$ & $-9.94$ \\

187 & $85.5031$ & $-69.1936$ & $4100$ & $0.09$ & $1055$ & $0.10$ & $--$ & --- & $8.68$ & $0.02$ & $-9.82$ \\

188 & $85.6608$ & $-69.1643$ & $2245$ & $0.14$ & $--$ & --- & $--$ & --- & $8.82$ & $0.02$ & $-9.68$ \\

189 & $85.7585$ & $-69.0972$ & $445$ & $0.05$ & $--$ & --- & $--$ & --- & $8.37$ & $0.03$ & $-10.13$ \\

195 & $81.9115$ & $-69.4793$ & $2550$ & $0.13$ & $--$ & --- & $--$ & --- & $8.22$ & $0.02$ & $-10.28$ \\

199 & $83.0805$ & $-67.5223$ & $3355$ & $0.16$ & $--$ & --- & $--$ & --- & $8.05$ & $0.03$ & $-10.45$ \\

200 & $83.8292$ & $-67.0387$ & $415$ & $0.19$ & $--$ & --- & $--$ & --- & $8.34$ & $0.03$ & $-10.16$ \\

205 & $83.2494$ & $-68.5986$ & $--$ & --- & $--$ & --- & $--$ & --- & $7.65$ & $0.03$ & $-10.85$ \\

208 & $80.4833$ & $-67.2127$ & $2665$ & $0.07$ & $--$ & --- & $--$ & --- & $8.20$ & $0.03$ & $-10.30$ \\

209 & $78.5745$ & $-67.3423$ & $1365$ & $0.05$ & $--$ & --- & $--$ & --- & $8.27$ & $0.02$ & $-10.23$ \\

210 & $79.9719$ & $-68.0677$ & $750$ & $0.09$ & $1625$ & $0.11$ & $--$ & --- & $7.20$ & $0.02$ & $-11.30$ \\

216 & $86.1350$ & $-70.6063$ & $580$ & $0.12$ & $2565$ & $0.13$ & $--$ & --- & $7.93$ & $0.03$ & $-10.57$ \\

217 & $83.0539$ & $-66.9883$ & $430$ & $0.14$ & $2570$ & $0.18$ & $--$ & --- & $8.42$ & $0.02$ & $-10.08$ \\

219 & $82.5861$ & $-66.8839$ & $3650$ & $0.17$ & $--$ & --- & $--$ & --- & $7.72$ & $0.03$ & $-10.78$ \\

220 & $83.1484$ & $-67.9192$ & $770$ & $0.28$ & $--$ & --- & $--$ & --- & $7.64$ & $0.02$ & $-10.86$ \\

223 & $85.6478$ & $-69.1467$ & $2100$ & $0.13$ & $--$ & --- & $--$ & --- & $7.71$ & $0.02$ & $-10.79$ \\

225 & $74.4305$ & $-70.1473$ & $955$ & $0.27$ & $--$ & --- & $--$ & --- & $7.32$ & $0.03$ & $-11.18$ \\

227 & $82.0619$ & $-66.5461$ & $445$ & $0.27$ & $--$ & --- & $--$ & --- & $8.84$ & $0.02$ & $-9.66$ \\

\hline

\end{tabular}
\end{table*}

\section{RESULTS AND DISCUSSION}
\label{sec:Results} 
\subsection{Period--Luminosity Relation}
\label{sec:P-Lsec} % used for referring to this section from elsewhere

\begin{figure*}
\centering
%\begin{subfigure}[b]{0.50\textwidth}
\includegraphics[width=1\linewidth]{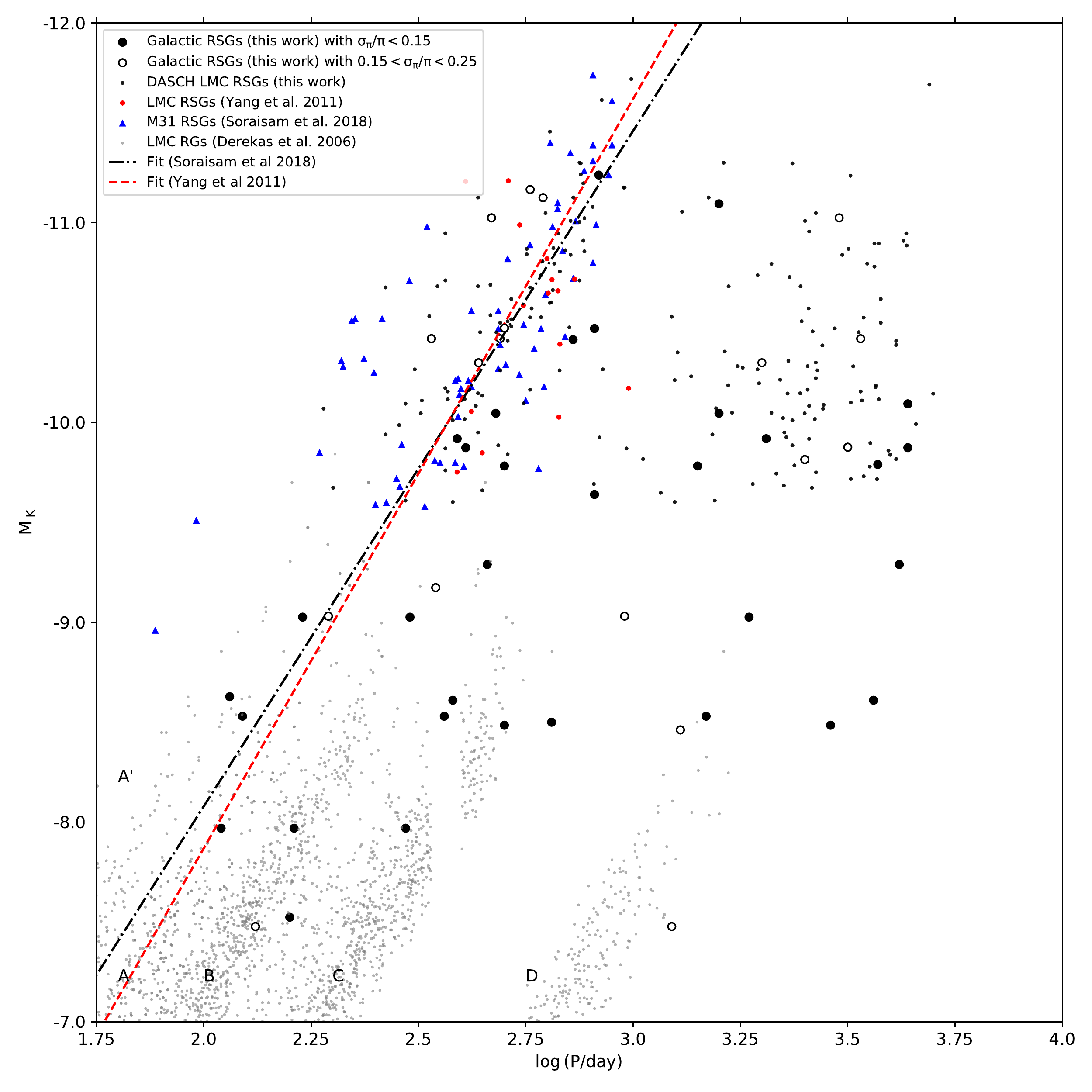}
\caption[]{Period--luminosity relation of our Galactic and LMC samples. Galactic RSGs with different fractional parallax uncertainty limits are shown as black filled ($15$\%) and empty circles ($25$\%). Best fit lines for the LMC and M31 are from \citet{YangJiang+2011} and \citet{Soraisam+2018}, respectively.}
\label{fig:PL-plots}
\end{figure*}

Figure {\ref{fig:PL-plots}} shows P--L diagram ($M_K$ versus log P) for our samples, with Galactic RSGs shown with two different thresholds on the parallax uncertainties (15\% and 25\%).
In addition to the RSGs analysed in this study, we show RSGs in M31 published by \citet{Soraisam+2018} and $14$ shorter periods in the LMC, published by \citet{YangJiang+2011} (see further discussion in Section\:{\ref{sec:RSGsinLMC}}). We also plot fitted lines from \citet{Soraisam+2018} and \citet{YangJiang+2011}. The agreement between the lines and our results is evident. \citet{Soraisam+2018} found that RSG P--L relation is consistent between the LMC, SMC, Galactic RSGs and the M33 (\citealt{YangJiang+2011}, \citealt{YangJiang+2012}, \citealt{Kiss++2006} and \citealt{Soraisam+2018}), despite the range of metallicities (\citealt{Ren+2019}).

\begin{figure}
\centering
\begin{subfigure}[a]{1\linewidth}
   \includegraphics[width=\linewidth]{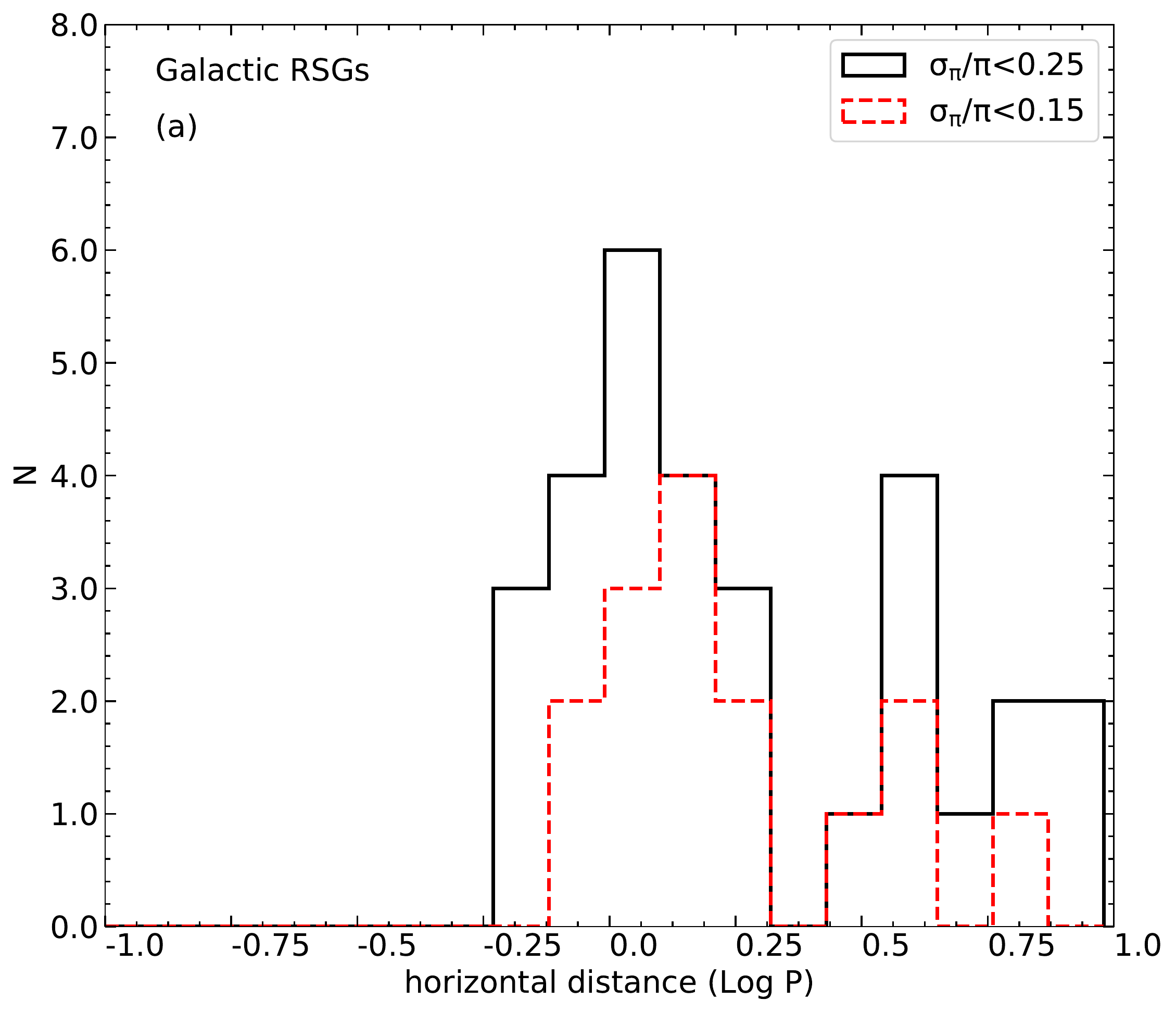}
\end{subfigure}

\begin{subfigure}[b]{1\linewidth}
   \includegraphics[width=\linewidth]{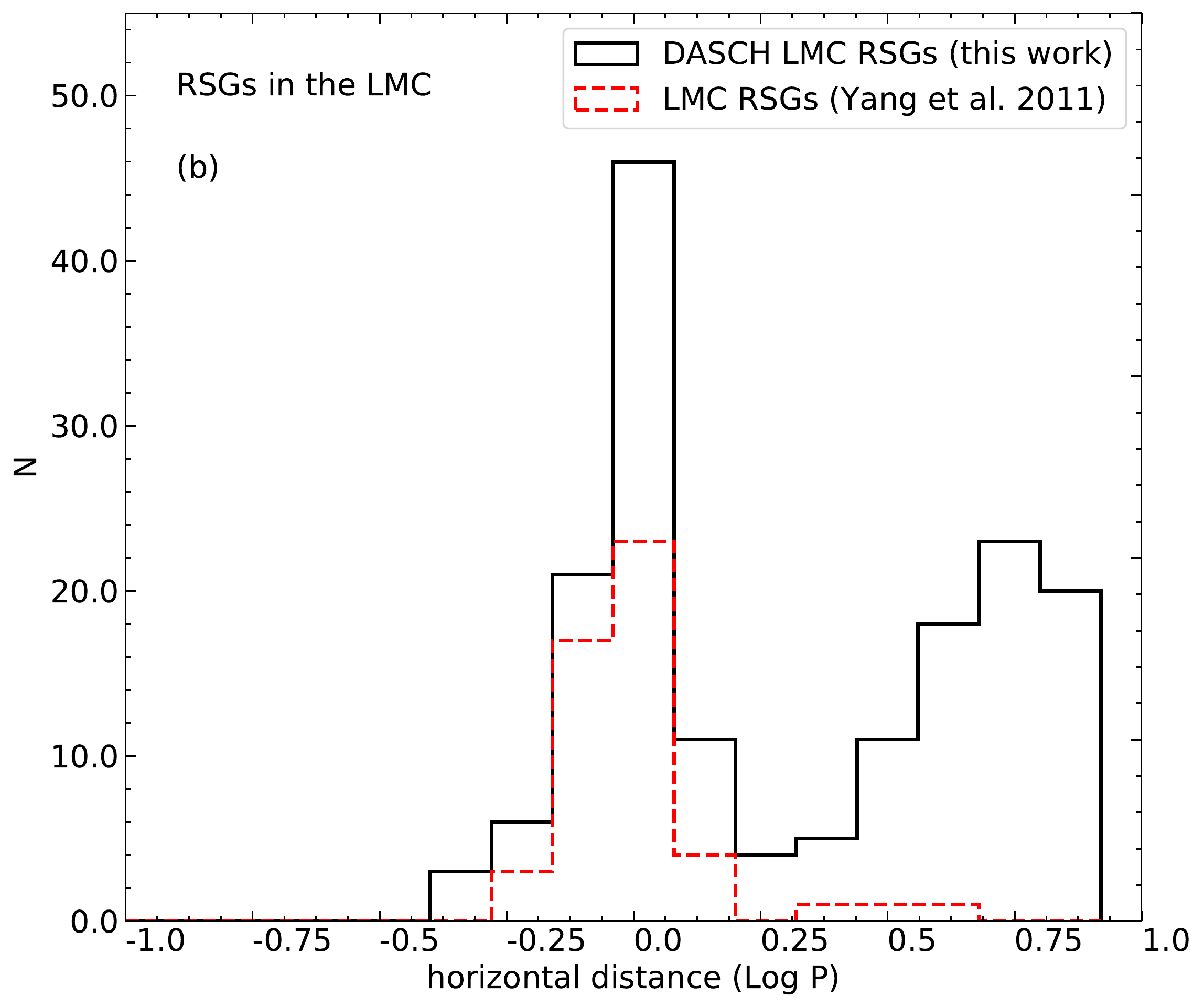}
\end{subfigure}
\caption[]{(a) Period distribution from the \citet{Soraisam+2018} fit line for our Galactic and LMC samples, as presented on Fig.\,{\ref{fig:PL-plots}}. (b) (RSGs in the LMC) Distribution of the detected DASCH periods is compared with shorter periods published in \citet{YangJiang+2011}. The two groups: (i) a presumed fundamental or low overtone sequence, and (ii) long secondary periods are distinct.}
\label{fig:hist}
\end{figure}

Figure\,{\ref{fig:hist}a} shows period distribution relative to the \citet{Soraisam+2018} fit line for our Galactic RSGs, with different fractional parallax uncertainty limits (15\% and 25\%). We remind the reader that the  uncertainties of the DR2 parallaxes need to be treated with caution (see Section\,{\ref{sec:DR2}} for further discussion).
The RSGs form two distinct groups: (i) a presumed fundamental or low overtone sequence, and (ii) long secondary periods with more scatter. The LSP scatter is discussed further in Section\:{\ref{sec:RGscomparison}}. We associate the shorter periods with the fundamental or low overtone modes of pulsation (\citealt{StothersLeung+1971}, \citealt{LiGing+1991}, \citealt{GuoLi+2002}, \citealt{Kiss++2006}) that are stochastically driven by convective motions in the envelope (\citealt{Schwarzschild+1975}, \citealt{Dalsgaard+2001}, \citealt{Bedding+2003}, \citealt{Kiss++2006}). 

The luminosity boundaries for identifying RSGs vary in the literature between $M_{\text{bol}}$=~$-5$\,mag (\citealt{Maeder+2000}) and $M_{\text{bol}}$\,=~$-7.0$\,mag (\citealt{Massey+LMC+2003}, \citealt{Wood+1983}) for the lower limits, to $M_{\text{bol}}$\,=~$-9$\,mag (\citealt{Maeder+2000}) and $M_{\text{bol}}$\,=~$-10.0$\,mag for the upper limits (\citealt{Humphreys+1986}). The upper limit corresponds to the Eddington luminosity, which indicates a stability boundary that prevents massive stars from evolving to cooler temperatures (\citealt{Humphreys+1986}, \citealt{dejager+1991}, \citealt{Lamers+1997}). It is a point where the radiation pressure can overcome gravity in the atmosphere of a star, making it unstable. This causes a significant mass loss in massive stars that are evolving to the right of the Hertzsprung--Russell diagram when cooling (\citealt{Lamers+2017}, \citealt{Levesque+2017}).

Assuming that $m_{\text{bol}}$\,$\approx$~$m_K$+$3$ (as per \citealt{Josselin+2000}), we set boundaries of our P--L diagrams between $M_K$\,$\approx$~$-7.0$\,mag for the faintest and $-12.0$\,mag for the brightest objects.
W~Cep and $\mu$~Cep, ($M_K$\,<~$-12.65$\,mag and $M_K$\,=~$-13.0$\,mag, respectively) along with SS~And, W~Tri and IS~Gem ($M_K$\,=~$-6.72$\,mag, $M_K$\,=~$-6.33$\,mag and $M_K$\,=~$-2.87$\,mag, respectively) fall outside our P--L diagrams.
 
Note that the histograms in Fig.\,{\ref{fig:hist}a} include Galactic RSGs that are brighter than $M_K$\,=~$-9.0$\,mag. This corresponds to $M_{\text{bol}}$\,=~$-6$\,mag. We adopted this cutoff for the presentation of the plot because it is the average of the two published lower limits of $M_{\text{bol}}$\,=~$-5$\,mag (\citealt{Maeder+2000}) and $M_{\text{bol}}$\,=~$-7.0$\,mag (\citealt{Massey+LMC+2003}, \citealt{Wood+1983}).

\subsection{Galactic RSGs}
\label{sec:RSGsinMW} % used for referring to this section from elsewhere

\begin{figure*}
\includegraphics[width=0.80\linewidth]{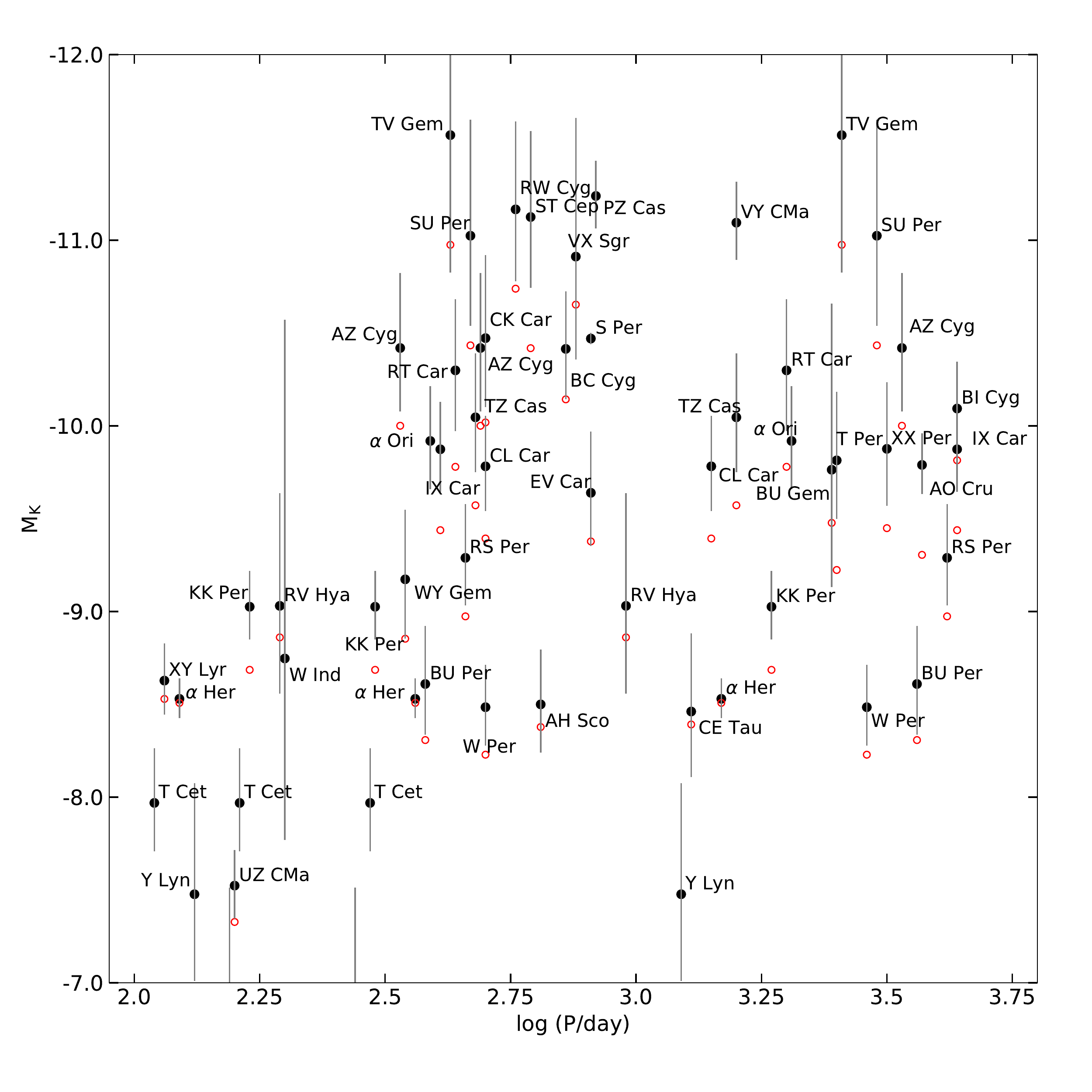}
\caption{Period--absolute $K$\,magnitude relations for our Galactic RSG sample, with fractional parallax uncertainty limit <\,$0.5$. $M_K$ uncertainties are shown as vertical lines. Red circles indicate shifted absolute magnitudes of the RSGs (RSGs with the DR2 parallaxes only), where a Gaia zero-point error of $0.1$\,mas has been adopted. See Section\,{\ref{sec:DR2}} for further discussion. W~Cep, $\mu$~Cep with their respective calculated absolute magnitudes of $M_K$\,<~$-12.65$\,mag and $M_K$\,=~$-12.97$\,mag are not shown, neither are SS~And, W~Tri, IS~Gem (calculated absolute $K$\,magnitudes $-6.72$, $-6.33$, $-2.87$\,mag, respectively). }
\label{fig:P-Lfind}
\end{figure*}

Figure\,{\ref{fig:P-Lfind}} presents period--$M_K$ relations for RSGs in our galactic sample. For convenience, the stars are labelled with their General Catalogue of Variable Stars (GCVS) names and error bars showing calculated $M_K$ uncertainties. Parallax uncertainty was the only factor included in the calculated $M_K$ uncertainties as majority of the $K$\,-band magnitude uncertainties were unavailable. 

 We found $10$ stars with relatively large $M_K$ uncertainties (>\,$1.0$\,mag; Y~Lyn, RV~Hya, SU~Per, VX~Sgr, SS~And, BU~Gem, NO~Aur, $\mu$~Cep, W~Ind, W~Cep) and it would be interesting to see their P--L relations with improved parallax measurements, anticipated in Gaia DR3.

Global astrometric satellites like Hipparcos and Gaia are able to measure absolute parallaxes, that is, without zero-point error, but this capability is susceptible to various instrumental effects which can lead to a small offset in the parallaxes (\citealt{DR2+astrometric+2018}). We investigated an impact of the zero-point offset of $-0.1$ mas  (\citealt{Khan+2019}) on the P--L relation of the Galactic RSGs. In Fig.\,{\ref{fig:P-Lfind}} we show that the adjusted parallaxes can significantly affect the absolute magnitudes of the most distant stars, with SU Per, RW Cyg, ST Cep, CL Car, RT Car and T Per shifted towards fainter values by $\sim$\,$0.5$ mag.

\begin{figure}
\includegraphics[width=\linewidth]{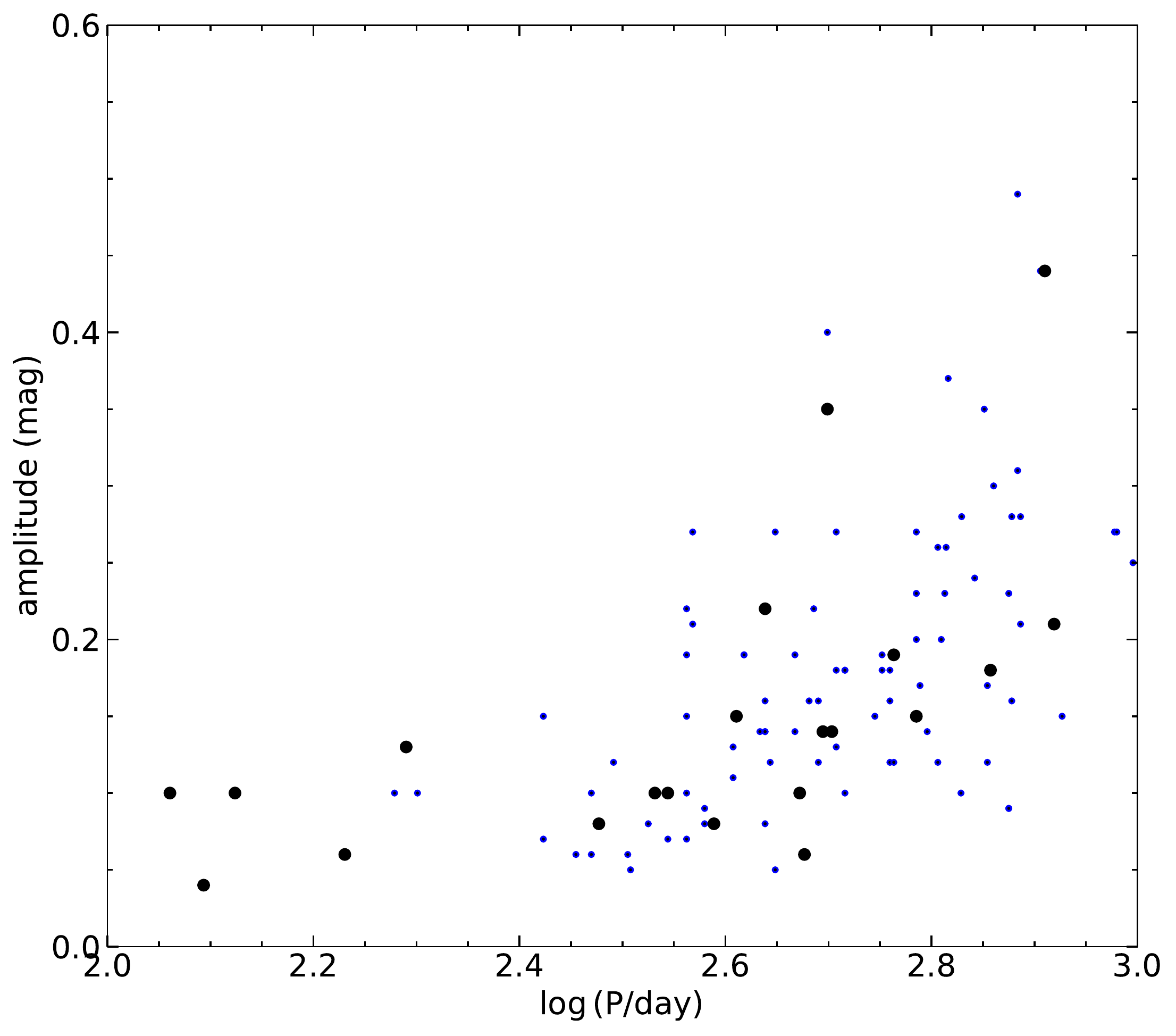}
\caption{The period--amplitude relationship of RSGs that are thought to be pulsating in the fundamental or low overtone modes. The LMC sample is shown as small blue circles, and Galactic RSGs are presented as large black circles. Note that our amplitudes are half of the peak-to peak amplitudes. The plot shows Galactic RSGs with $\sigma_\pi$/$\pi$\,=~$0.25$.}
\label{fig:P-Samp}
\end{figure}

Overall, the indicated periods of ASAS and AAVSO data agreed well for majority of the sample. We favored higher quality lightcurves where they indicated different pulsational frequencies (seven stars). Among those, we found three objects with more reliable data from AAVSO (T~Per, AO~Cru, WY~Gem) and four from ASAS (TZ~Cas, EV~Car, XY~Lyr, SS~And).
Similarly to \citet{YangJiang+2011}(figure 10), our determined periods show positive correlation with the amplitudes (Fig.\,{\ref{fig:P-Samp}}), which is expected for solar--like variations (\citealt{Bedding+1995}).

As concluded by \citet{Kiss++2006}, the periods of RSGs can be divided into two groups, pulsations with periods of $300$--$1000$\,d and LSPs with periods of a few thousand days.
In total, we found $40$ shorter periods in $25$ stars and $23$ LSPs in $23$ stars. Some lightcurves were noisy and their periodicities could not be determined. 

\subsection{RSGs in the LMC}
\label{sec:RSGsinLMC} % used for referring to this section from elsewhere

Similarly to our Galactic sample, RSGs in the LMC show complex light variations on time scales that range from months to several years. We were able to observe two types of variation\,--\,pulsations similar to oscillations in other types of stars and a long secondary period, unique to red giants and supergiants. 

The P--L relation we have found agrees well with previous works. The fraction of stars for which we could detect periods ($83$\%) is significantly greater than the $51$\% found by \citet{YangJiang+2011}. We explain this difference by the fact that the DASCH observations cover three to six times longer in time. We also found LSPs in $94$ stars (or $55$\%).

In Figure\,{\ref{fig:hist}b} we show a good agreement between distributions of the DASCH periods ($92$ shorter and $95$ long secondary periods) and the $47$ shorter periods published in \citet{YangJiang+2011}).
We note that we do not show LSPs, published by \citealt{YangJiang+2011} because they were based on only $3000$\,d of data (mostly ASAS), which we consider too short for a reliable detection of LSPs. Thus, our P--L diagram in Fig.\,{\ref{fig:PL-plots}} shows only shorter periods, published in \citet{YangJiang+2011} and only those for which we could not identify periods from the DASCH data ($14$ objects in total). They lay along the \citet{Soraisam+2018} best fit line, in a good agreement with the detected shorted periods in this study.

\subsection{Comparing red supergiants to red giants}
\label{sec:RGscomparison} % used for referring to this section from elsewhere
In Figure\,{\ref{fig:P-Lfind}} we show the corresponding LMC sequences from the Massive Compact Halo Object (MACHO) survey data, analysed by \citet{Derekas+2006}, for a comparison. The sequences are labelled C (fundamental), A$'$, A, B (overtones) and D (LSPs) following the naming conventions by \citet{Wood+1999}. The sequence A$'$ comprise the shortest periods and smallest amplitudes (\citealt{Soszynski+2004}). LSPs are common for RSGs and RGs but the origin of this phenomenon has been a long-standing unknown (\citealt{Stothers+2010}).
The RSGs are clearly not in line with the relatively tight sequences of the red giants, as might have been expected. The RSGs are clustered in two groups: pulsations (presumably fundamental and low overtones) and LSPs (around extended sequence D of the RGs), with the LSPs group being much more dispersed. In general, the observed scatter in RSGs may be a result of the fact that the evolutionary tracks of different masses overlap in luminosity, which in turn can affect their periods (\citealt{Soraisam+2018}). Models show that the overlap can be even stronger, depending on how other underlying processes such as convection, binarity and mass loss (which are currently not entirely understood) are treated (\citealt{Levesque+2017}). These add uncertainty to any potential extragalactic distance estimates based on the P--L relations of RSGs. 

An interesting feature of the presented P--L relations is a lack of stars between RSGs and RGs (between $M_K$$\approx$\,$-8.5$ and $M_K$$\approx$\,$-9.5$\,mag). Future study of objects that occupy this gap should reveal what transition in the P--L diagram they form between RSG and RGs.

\subsection{RSGs in M31}
\label{sec:M31comparison} % used for referring to this section from elsewhere

\citet{Soraisam+2018} studied RSGs in M31 (Andromeda) and found that the P--L relation of the RSGs  agree well across the nearest galaxies (SMC, LMC, Milky Way, M31). They compared the analyzed P--L relation with the theoretical one based on MESA models and found their two groups of shorter periods to pulsate in the fundamental radial and low overtone modes.

The majority of the shorter period M31 RSGs (shown as blue triangles in Fig.\,{\ref{fig:PL-plots}}) overlap with our Galactic and the LMC shorter periods, around the extended sequence A of the RGs. In general, the agreement is good, with a few stars in M31 that seem to pulsate with shorter periods than expected for their luminosities. \citet{Soraisam+2018} suggested that these stars may have different masses but overlapping luminosities. No LSPs have been published in the Andromeda galaxy study as it is based on the five-year survey, which is too short for an accurate detection of LSPs.
 
\section{Conclusions}

\label{sec:Conclusion} % used for referring to this section from elsewhere

Using long-term lightcurves from several campaigns we analysed over $220$ RSGs in the LMC and the Milky Way and studied their main pulsational characteristics. Parallaxes from Gaia Data Release 2 allowed us to tighten the P--L relations of our Galactic sample, where we found $40$ shorter and $23$ longer periods. Our LMC sample contains 142 stars, with most having a usable observation time of approximately $50$ years. Among those, we found $92$ shorter and $95$ longer periods.

We found that the P-$M_K$ relations agree well with the literature (\citealt{YangJiang+2011}, \citealt{Soraisam+2018}).
When compared to the red giants, it is clear that the RSGs do not follow the same sequences. Periods of RSGs form two groups: (i) a pronounced group on the P--L diagram (Fig.\,{\ref{fig:PL-plots}}) with periods of $300$--$1000$\,d, and (ii) the LSP group, with periods between $1000$ and $8000$\,d, that is much more dispersed. We considered an impact of the Gaia zero-point shift in parallaxes on the P--L relations of distant RSGs to be significant.

It is clear that pulsations following a P--L relation are present in most RSGs in the Local Group and that this relation does not depend on the metallicity \citep{Ren+2019}. In order to consider RSGs ``standard candles'' (as suggested by \citealt{Glass+1979}), factors like the abundance of irregular variables, mass loss, dust production and the additional sources of long-period variability need to be further explored. Each of these mechanisms can contribute to changes in the apparent magnitude of RSGs, causing periodic or stochastic fluctuations in their lightcurves, and result in a further dispersion of their P--L relations.

Without a theoretical basis, further investigation of origins of the LSPs and variability of RSGs in general, is difficult.  Long-term photometric monitoring is one of the main challenges in studying pulsations of RGSs, and there may be many years before we expect better lightcurves. More precise distance measurements from future Gaia data releases may present opportunities for future work.

\section*{Acknowledgements}

We thank Grzegorz Pojmanski for his valuable input and for providing the ASAS photometry of Galactic RSGs.
We acknowledge with thanks the variable star observers of the AAVSO whose many decades of observations were used in this research.
The DASCH project at Harvard is grateful for partial support from NSF grants AST-0407380, AST-0909073, and AST-1313370. We thank the anonymous referee for their detailed comments,
which improved the paper significantly. This research has made use of the VizieR and SIMBAD databases, operated at CDS, Strasbourg, France and the NASA ADS Abstract Service. This work has made use of data from the European Space Agency (ESA) mission {\it Gaia} (\url{https://www.cosmos.esa.int/gaia}), processed by the {\it Gaia} Data Processing and Analysis Consortium (DPAC,\url{https://www.cosmos.esa.int/web/gaia/dpac/consortium}). Funding for the DPAC has been provided by national institutions, in particular the institutions participating in the {\it Gaia} Multilateral Agreement. Dougal Dobie is supported by an Australian Government Research Training Program Scholarship. L\'aszl\'o L. Kiss has been supported by the LP2018-7 Lend\"ulet grant of the Hungarian Academy of Sciences and the K-115709 and the GINOP-2.3.2-15-2016-00003 grants of the National Research, Development and Innovation Office (NKFIH, Hungary).

%%%%%%%%%%%%%%%%%%%%%%%%%%%%%%%%%%%%%%%%%%%%%%%%%%

%%%%%%%%%%%%%%%%%%%% REFERENCES %%%%%%%%%%%%%%%%%%

% The best way to enter references is to use BibTeX:

\bibliographystyle{mnras}
\bibliography{RSGsPL.bib} % if your bibtex file is called example.bib

\appendix
\section{Light curves and power spectra}

Supplementary material is available online.
%%%%%%%%%%%%%%%%%%%%%%%%%%%%%%%%%%%%%%%%%%%%%%%%%%

%%%%%%%%%%%%%%%%% APPENDICES %%%%%%%%%%%%%%%%%%%%%

%%%%%%%%%%%%%%%%%%%%%%%%%%%%%%%%%%%%%%%%%%%%%%%%%%

% Don't change these lines
\bsp	% typesetting comment
\label{lastpage}
\end{document}